\documentclass[
 aps,
 pra,
 amsmath,amssymb,
 superscriptaddress,
 reprint
]{revtex4-1}

\usepackage{graphicx}
\usepackage{dcolumn}
\usepackage{bm}

\usepackage[utf8]{inputenc}
\usepackage[T1]{fontenc}
\usepackage{mathptmx}
\usepackage{etoolbox}

\usepackage{braket}
\usepackage{color}

\begin{document}

\title{
High-performance multiqubit system with double-transmon couplers: 
Toward scalable superconducting quantum computers
}
\author{Kentaro Kubo}
\email{kentaro3.kubo@toshiba.co.jp}
\affiliation{
Frontier Research Laboratory, Corporate Research \& Development Center, Toshiba Corporation, 1, Komukai-Toshiba-cho, Saiwai-ku, Kawasaki 212-8582, Japan.
}
\author{Yinghao Ho}
\affiliation{
Frontier Research Laboratory, Corporate Research \& Development Center, Toshiba Corporation, 1, Komukai-Toshiba-cho, Saiwai-ku, Kawasaki 212-8582, Japan.
}
\author{Hayato Goto}
\affiliation{
Frontier Research Laboratory, Corporate Research \& Development Center, Toshiba Corporation, 1, Komukai-Toshiba-cho, Saiwai-ku, Kawasaki 212-8582, Japan.
}
\affiliation{RIKEN Center for Quantum Computing (RQC), Wako, Saitama 351-0198, Japan
}

\date{\today}

\begin{abstract}
Tunable couplers in superconducting quantum computers 
have enabled fast and accurate two-qubit gates, 
with reported high fidelities over 99\% 
in various architectures and gate implementation schemes.
However,
there are few tunable couplers 
whose performance in multi-qubit systems is clarified, 
except for the most widely used one: single-transmon coupler (STC). 
Achieving similar accuracy to 
isolated two-qubit systems 
remains challenging due to various undesirable couplings but is necessary for scalability.
In this work, 
we numerically analyze a system of three fixed-frequency qubits 
coupled via two double-transmon couplers (DTCs)
where nearest-neighbor qubits are highly detuned
and also next nearest-neighbor ones are nearly resonant.
The DTC is a recently proposed tunable coupler, 
which consists of two fixed-frequency transmons 
coupled through a common loop with an additional Josephson junction. 
We find that the DTC can not only 
reduce undesired residual couplings sufficiently, 
as well as in isolated two-qubits systems, 
but also enables implementations 
of 30-ns CZ gates 
and individual and simultaneous 10-ns $\pi/2$ pulses 
with fidelities over 99.99\%. 
For comparison,
we also investigate the system 
where the DTCs are replaced by the STCs. 
The results show that the DTC outperforms the STC 
in terms of both residual coupling suppression 
and gate accuracy 
in the above systems. 
From these results, 
we expect that the DTC architecture is promising 
for realizing high-performance, scalable superconducting quantum computers.
\end{abstract}

\maketitle

\section{Introduction}
The advent of tunable couplers 
has dramatically improved the gate performance of 
superconducting quantum processors
\cite{spremacy1,spremacy2,yan2018tunable}. 
They can substantially suppress 
undesired residual couplings 
by adjusting external parameters such as magnetic flux, 
and also quickly strengthen the couplings to implement 
fast two-qubit gates.

Although various types of tunable couplers have been proposed
\cite{yan2018tunable,fluxqubit1,fluxqubit2,fluxonium1,fluxonium2,fluxonium3,longdistance,
doubleresonator,gmon1,gmon2,inductive1_parametric1,inductive2_parametric2,floatingSTC,CSfluxqubit},
so far the most standard, 
widely used, and well-understood one is the single-transmon coupler (STC)\cite{yan2018tunable}. 
The STC has a simple structure consisting 
of a frequency-tunable transmon and capacitor between qubits\cite{yan2018tunable}. 
Several groups have reported fast and accurate 
(various types of) two-qubit gate implementations 
using an STC in isolated two-qubit systems
\cite{PRXSTC, RiggettiSTC,STCChinese, ETH, IBMSTC}. 
Furthermore, it has been confirmed that 
the functionality of the STC scales to multi-qubit systems\cite{spremacy1,spremacy2,TheoriticalSTCcrosstalk2022,YanSTCCZ,sycamore2,chinese2}. 
However, 
it is known that the 
reduction of residual couplings
for highly detuned qubits (larger than qubit anharmonicities)
is challenging for the STC\cite{YanSTCCZ,STCChinese, ETH, IBMSTC}. 
For example, it has been reported 
that there is
the so-called residual $ZZ$ coupling of -80 kHz to -60 kHz
for about 360 MHz detuned qubits\cite{ETH}. 
Because of this drawback, 
a standard architecture 
uses the STC together with frequency-tunable qubits with a small detuning 
(less than qubit anharmonicities)
\cite{spremacy1,spremacy2,yan2018tunable,PRXSTC, RiggettiSTC,sycamore2,chinese2}.
However, 
such nearly resonant qubits tend to suffer from
frequency crowding and microwave crosstalk compared to highly detuned ones.
Furthermore, 
tunable qubits 
are prone to decoherence compared with fixed-frequency ones\cite{Koch2007,krantz2019}.

\begin{figure}[t]
\includegraphics[width=0.48\textwidth]{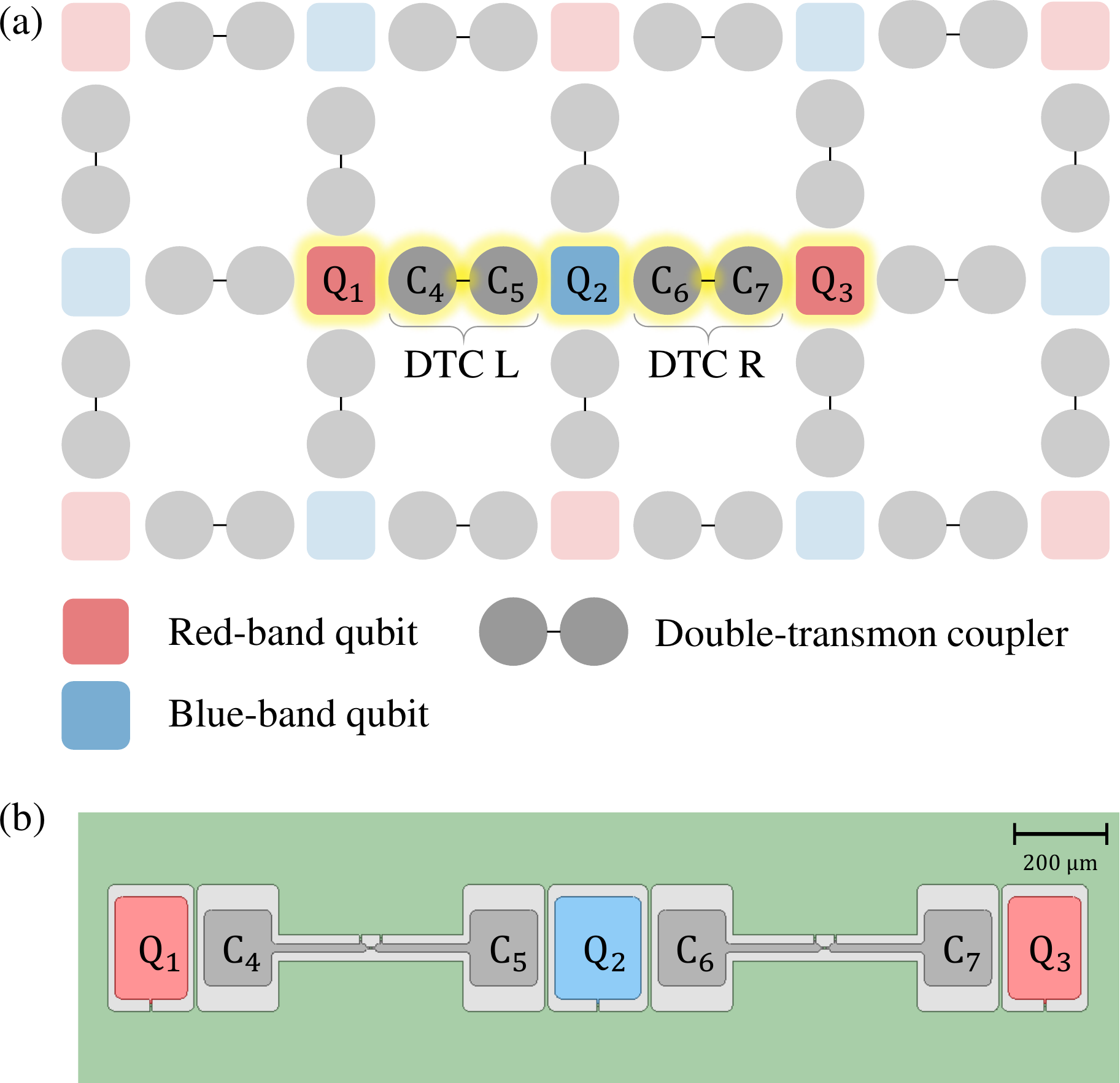}
\caption{
(a) 
A typical structure of a multiqubit system with the DTCs. 
Squares and circles represent qubits and coupler transmons, respectively. 
The colors of qubits (red or blue) indicate 
which frequency bands (low or high) they belong to. 
DTCs are represented by two coupler transmons connected via a thin line. 
(b) A typical layout of the highlighted part of (a) 
corresponding to the three-qubit system that we study in this work.
Its schematic circuit diagram is shown in Fig.\ \ref{fig:DTC}(a).
Its capacitances estimated by electromagnetic simulator ANSYS Q3D\cite{Ansys}
are summarized in Table \ref{tab:paraDTC}.
}
\label{fig:redblue}
\end{figure} 
\begin{figure*}[t]
\centering
\includegraphics[width=0.93\textwidth]{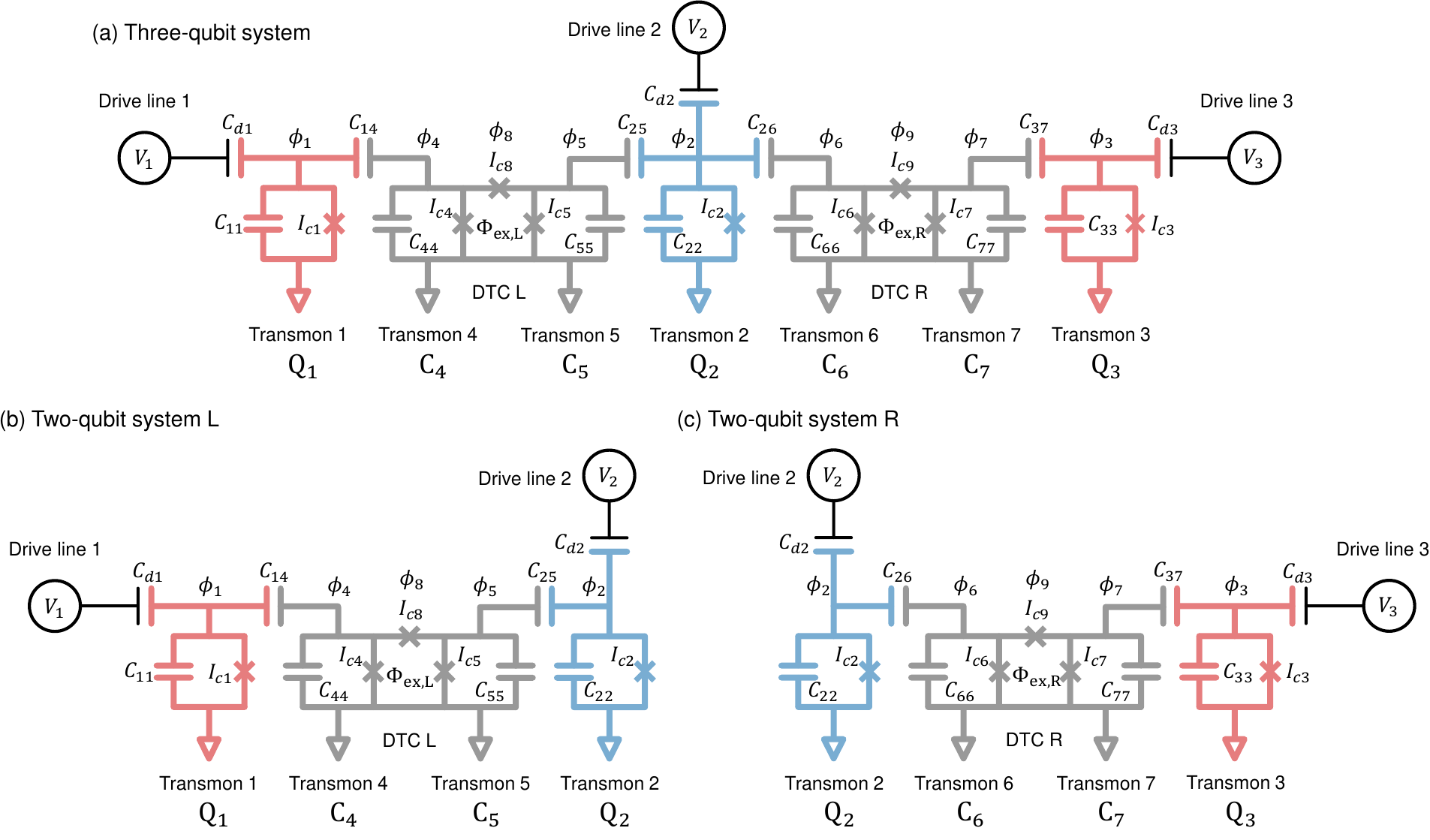}
\caption{
(a) Circuit diagram of the three qubit system 
corresponding to the highlighted part of Fig.\ \ref{fig:redblue}. 
(b) Circuit diagram of the two-qubit system L, 
which is the 
subsystem of (a). 
(c) Similar diagram to (b) for the right side of (a). 
The colors of ${\rm Q}_i$ (red or blue) indicate 
which frequency bands (low or high) they belong to.  
}
\label{fig:DTC}
\end{figure*} 

To overcome this disadvantage of the STC, 
a new kind of tunable coupler named the double-transmon coupler (DTC) 
has recently been proposed\cite{goto2022DTC}, 
numerically investigated\cite{KG2023,Airforce2023}, 
and experimentally realized\cite{RikenToshiba2024}.
The DTC consists of two fixed-frequency transmons coupled through a common loop with an addittional Josephson junction\cite{goto2022DTC,KG2023,Airforce2023,RikenToshiba2024}. 
It has been 
numerically\cite{goto2022DTC,KG2023} and experimentally\cite{RikenToshiba2024}
demonstrated that
the DTC can make 
the residual $ZZ$ coupling strength completely zero
or negligibly small 
even for highly detuned qubits, which has not been achieved with STCs. 
Numerical gate simulations using rigorous superconducting-circuit models 
without decoherence effect 
have shown that 
a fast CZ gate using a tunable longitudinal coupling\cite{goto2022DTC,KG2023}
and a fast $\sqrt{\rm iSWAP}$ gate using a parametric transverse coupling\cite{KG2023} 
with fidelities over 99.99\% can be implemented 
for highly detuned fixed-frequency transmon qubits. 
Furthermore, 
in the experimental work\cite{RikenToshiba2024}, 
a CZ-gate fidelity of 99.92$\pm$0.01\% has been realized stably 
during 12-hour measurement.
This is the highest level of two-qubit gate fidelity 
among superconducting quantum computers ever reported. 
It has also been reported in Ref.\ \cite{RikenToshiba2024} that 
the coherence times of qubits coupled via a DTC 
are also at the highest level as transmons 
($T_1=228.6, 205.3\ \mu$s and $T_2^{\rm E}=358.9, 129.8\ \mu$s at the idle point).
These results suggest that 
the degradation of the coherence time 
due to the noise channels introduced by a DTC 
may be rather smaller than or 
at least comparable to
that of conventional tunable couplers. 

However,
all studies of the DTC reported so far 
have been done only in isolated two-qubit systems\cite{goto2022DTC,KG2023,Airforce2023,RikenToshiba2024}
and 
therefore it has not been clear whether 
the above abilities of the DTC can be 
maintained in multi-qubit systems like Fig.\ \ref{fig:redblue}(a). 
In other words, 
the scalability of the DTC architecture has not been understood well, 
unlike the other architectures including the STC\cite{spremacy1,spremacy2,TheoriticalSTCcrosstalk2022,YanSTCCZ,sycamore2,chinese2,
crosstalk1,crosstalk2,crosstalk3,crosstalk4,crosstalk5}. 
Since there are additional error sources in multi-qubit systems, 
such as unwanted interactions 
between nonadjacent qubits, 
between a qubit and a nonadjacent coupler, 
and between couplers, 
the detailed study on multi-qubit systems with multiple DTCs is highly desirable.

In this paper, 
we numerically study the system with three qubits coupled via two DTCs as a minimal model
for the above purpose
[see the highlighted part in Fig.\ \ref{fig:redblue}(a)].
As a result, we find that $ZZ$ coupling between nearest-neighbor (NN) qubits can be reduced 
as in isolated two-qubit systems. 
Furthermore, 
we show that 
the $ZZ$ coupling between next-nearest-neighbor (NNN) qubits and $ZZZ$ coupling, 
which do not exist in isolated two-qubit systems, 
can also be suppressed down to about 1 kHz.
We also numerically demonstrate that 
30-ns CZ gates  
and 
individual and simultaneous 10-ns $\pi$/2 pulses 
can be implemented with fidelities over 99.99\%.
These results imply that the DTC 
works well even in multi-qubit systems.
Moreover, as a comparison of the above results, 
we also evaluate the performance 
of the system
where the two DTCs are replaced by the two STCs. 
The parameters of STCs have been set based on 
the experiment reporting a high-performance CZ gate\cite{ETH}.
The results show that the DTC outperforms the STC 
in terms of both residual coupling suppression 
and gate accuracy 
in the above systems. 
Based on our findings, 
the main reason for the difference lies in the fact 
that the STC architecture typically exhibits a larger stray coupling 
between the NNN qubits compared to the DTC architecture (see Sec.\ \ref{sec:Discussion}).

This paper is organized as follows. 
In Sec.\ \ref{sec:Model}, 
we introduce 
a theoretical model of the above three-qubit system in Fig.\ \ref{fig:redblue}
and show the present parameter setting. 
In Sec.\ \ref{sec:IdlingPoints}, 
we show the numerical results of the $ZZ$ and $ZZZ$ couplings 
and also  
two-qubit and single-qubit gates (CZ gates 
and individual and simultaneous $\pi$/2 pulses, respectively). 
For the evaluation of gate performance, we use 
the average gate fidelities \cite{fidelity,kueng2016comparing} for the three-qubit system, 
which can include the effects of spectator errors\cite{TheoriticalSTCcrosstalk2022,spectator1,spectator2,spectator3,spectator4,spectator5,spectator6,spectator7}.
In Sec.\ \ref{sec:STC}, 
for comparison,
we also analyze the system 
where the DTCs are replaced by the STCs. 
In Sec.\ \ref{sec:Discussion}, 
we discuss 
the difference between the results of  the DTCs and the STCs. 
Finally, 
we summarize our work in Sec.\ \ref{sec:summary}.

\section{Model}
\label{sec:Model}
\subsection{Circuit}
We consider the three-qubit system shown by the highlighted part in Fig.\ \ref{fig:redblue}(a). 
A typical device layout 
and a schematic circuit diagram are shown in Fig.\ \ref{fig:redblue}(b) and Fig.\ \ref{fig:DTC}(a), respectively.
We also introduce 
the isolated two-qubit subsystems L and R shown in Figs.\ \ref{fig:DTC}(b) and \ref{fig:DTC}(c), 
respectively, 
corresponding to the left and right sides of the three-qubit system.

The three-qubit system consists of seven transmons. 
Transmons 1, 2, and 3 in Fig.\ \ref{fig:DTC}(a) 
are fixed-frequency qubits ($\rm{Q_1}$, $\rm{Q_2}$, and $\rm{Q_3}$). 
Neighboring two-qubit pairs, 
$(\rm{Q_1}, \rm{Q_2})$ and $(\rm{Q_2}, \rm{Q_3})$, 
are coupled via DTCs L and R, respectively. 
DTC L (R) consists of two fixed-frequency transmons, 
$\rm{C_4}$ and $\rm{C_5}$ ($\rm{C_6}$ and $\rm{C_7}$), 
coupled through a common loop with an additional Josephson junction,  
the critical current of which, $I_{c8(9)}$, is smaller than that of the transmons, 
$I_{ci}$ ($i\in\{1,2,3,4,5,6,7\}$).
In the loop of DTC $\mu$ ($\mu\in\{\rm L, \rm R\}$), 
the external magnetic flux $\Phi_{{\rm ex},\mu}$ is applied.
Each qubit ${\rm Q}_i$ is coupled to a drive line, 
where voltage $V_i$ is applied 
via a capacitor $C_{di}$ .
We assume that $C_{di}$ is negligible with respect to the other capacitances $C_{ij}$, 
where $C_{ii}$ is a capacitance between transmon $i$ and the ground, 
and $C_{ij}$ $(i\neq j)$ is the one between transmons $i$ and $j$. 

\subsection{Hamiltonian of the superconducting circuit model}
\label{sec:Hamiltonian}
We assume $V_i=0$ except for 
single-qubit gates in Sec.\ \ref{sec:DTCsingle}. 
Then, 
the Hamiltonian of the rigorous circuit model of this system 
is written as follows 
(see Appendix \ref{sec:derivation} for the derivation): 
\begin{align}
	\hat{H} &= 4\hbar\hat{\mathbf{n}}^{\rm T} W \hat{\mathbf{n}} 
	+ 
	\left(
	\dot{\Theta}_{\rm ex,L}
	\mathbf{t}_{\rm DL}^{\rm T}
	+ \dot{\Theta}_{\rm ex,R}
	\mathbf{t}_{\rm DR}^{\rm T}
	\right)
	\hbar W\hat{\mathbf{n}}
	+ \hat{U}, 
	\label{eq:Hamiltonian}
\end{align}
where 
\begin{align}
	\hat{U}
	&= -\sum_{i=1}^7 \hbar\omega_{J_i}\cos\hat{\varphi}_i \nonumber\\
		&\ \ \ \ -\hbar\omega_{J8}\cos(\hat{\varphi}_{5}-\hat{\varphi}_4-\Theta_{\rm ex,L})  \nonumber\\
		&\ \ \ \ -\hbar\omega_{J9}\cos(\hat{\varphi}_{7}-\hat{\varphi}_6-\Theta_{\rm ex,R}), \\
\mathbf{t}_{\rm DL}^{\rm T}
&=
\frac{1}{\omega_{C_{45}}}
\begin{pmatrix}
	  0 & 0 & 0 & -1 & 1 &0 &0\\
\end{pmatrix}, 	
\label{eq:tDL}\\
\mathbf{t}_{\rm DR}^{\rm T}
&=
\frac{1}{\omega_{C_{67}}}
\begin{pmatrix}
	  0 & 0 & 0 &0 &0 & -1 & 1\\
\end{pmatrix},  
\label{eq:tDR}
\end{align}
$\hbar$ is the reduced Planck constant, 
$\Theta_{\rm ex, \mu}=\Phi_{\rm ex,\mu}/\phi_0$ is an angle defined 
with the external flux $\Phi_{\rm ex, \mu}$, 
$\phi_0=\hbar/(2e)$ is the reduced flux quantum, 
$\hbar W = e^2M^{-1}/2$ 
with a capacitor matrix $M$ ($M_{ii}=\sum_{j=1}^7C_{ij}$ and $M_{ij}=-C_{ij}$ for $i\neq j$), 
and $\hbar \omega_{C_{ij}}=e^2/(2C_{ij})$ (for $i\neq j$), with the elementary charge $e$. 
Operators $\hat{n}_i$,  $\hat{\varphi}_{i}$, and $\hbar\omega_{J_i}=\phi_0 I_{ci}$ are, respectively, 
the Cooper-pair number operator, 
the phase difference operator, 
and the Josephson energy for the $i$th Josephson junction. 
Operators $\hat{n}_i$ and $\hat{\varphi}_{i}$ satisfy 
the canonical commutation relation $[\hat{\varphi}_{i}, \hat{n}_j]=i\delta_{i,j}$.

In numerical simulations of superconducting quantum computers, 
effective models derived by approximating 
the rigorous circuit ones are widely used \cite{yan2018tunable,PRXSTC, RiggettiSTC,STCChinese, 
IBMSTC,YanSTCCZ,TheoriticalSTCcrosstalk2022}.
The reason for this is that these models are intuitive and computationally light.
However, these effective models 
may lead to inaccurate results compared to rigorous ones like the above. 
Therefore, in this study, 
we use the above rigorous superconducting circuit model without approximations, 
focusing on demonstrating the performance in an ideal situation in the absence of decoherence 
as rigorously as possible. 
In the case of a two-qubit system, 
it has already been reported that such simulation results 
are in excellent agreement with experimental ones\cite{RikenToshiba2024}.

The Hamiltonian in Eq.\ (\ref{eq:Hamiltonian}) 
is represented by a $(2N+1)^7\times(2N+1)^7$ matrix,  
where $N$ is a cutoff for the Cooper-pair number (see Appendix \ref{sec:MatrixRepresentation}).
In this work, 
we choose $N=10$ so that the energies converge sufficiently.
The matrix size of the Hamiltonian of our three-qubit system $(1801088541\times1801088541)$ 
is 85766121 times larger than 
the one in the two-qubit systems studed in Refs.\ \cite{goto2022DTC,KG2023} $(194481\times194481)$. 
Note that calculations of this size are too heavy to perform 
in a naive manner, e.g., by directly using QuTip\cite{qutip1,qutip2}. 
Due to these difficulties, 
there have been no studies 
that conducted gate simulations for multi-qubit systems 
coupled via tunable coupler
using rigorous circuit models, to the best of our knowledge. 
We overcome this difficulty introducing the dimension reduction technique in Appendix \ref{sec:simulation}. 
We have confirmed that 
calculated energies and gate fidelities converge 
with errors of the orders of sub-kHz and $10^{-5}$, respectively. 
Therefore, 
in this work 
we evaluate residual couplings up to 1-kHz rounding off the sub-kHz fractions
and the gate fidelities up to 4-digit precision rounding off the fifth decimal place.

The eigenfrequencies of the three-qubit system 
and corresponding eigenstates are denoted by
$\omega_{Q_1,Q_2,Q_3,C_4,C_5,C_6,C_7}$ and $\ket{Q_1,Q_2,Q_3,C_4,C_5,C_6,C_7}$ 
($Q_i, C_i \in {0,1,2,\cdots}$), respectively. 
Since we are mainly interested in the qubit subspace, 
we also use notations $\omega_{Q_1,Q_2,Q_3}\equiv\omega_{Q_1,Q_2,Q_3,0,0,0,0}$ and $\ket{Q_1,Q_2,Q_3}\equiv\ket{Q_1,Q_2,Q_3,0,0,0,0}$. 
Hereafter, 
we set $\omega_{0,0,0}$ to 0. 

\subsection{Parameter setting}
\label{sec:ParameterSetting}
\begin{table}
\label{tab:parametersDTC}
\caption{
Parameter setting for the DTC circuits in Fig.\ \ref{fig:DTC}. 
Bold values are design values. 
The others are calculated using them.}
\begin{minipage}{0.23\textwidth}
\vspace{-0.75cm}
\begin{ruledtabular}
\begin{tabular}{ll}
	$\omega_{1}/2\pi$ (GHz) & {\bf 5.0} \\ 
	$\omega_{2}/2\pi$ (GHz)  & {\bf 5.5} \\ 
   	$\omega_{3}/2\pi$ (GHz) & {\bf 5.01} \\ 
  	$\omega_{4}/2\pi$ (GHz) & {\bf 7.3} \\
  	$\omega_{5}/2\pi$ (GHz) & {\bf 7.3} \\
  	$\omega_{6}/2\pi$ (GHz) & {\bf 7.3} \\
  	$\omega_{7}/2\pi$ (GHz) & {\bf 7.3} \\
	$C_{ii}$ (fF) & {\bf 80.0} \\
	$C_{12}, C_{23}$ (fF)  & {\bf 0.05} \\
	$C_{13}$ (fF) & {\bf 0.003} \\
 	$C_{14},C_{25},C_{26},C_{37}$ (fF) & {\bf 8.0} \\
 	$C_{15},C_{24},C_{27},C_{36}$ (fF) & {\bf 0.1} \\
 	$C_{16},C_{35}$ (fF) & {\bf 0.02} \\
 	$C_{17},C_{34}$ (fF) & {\bf 0.006} \\
	$C_{45},C_{67}$ (fF) & {\bf 1.0} \\
 	$C_{46},C_{57}$ (fF) & {\bf 0.05} \\
	$C_{47}$ (fF) & {\bf 0.01} \\
 	$C_{56}$ (fF) & {\bf 1.0} \\
	$I_{c1}$ (nA) & 31.0 \\
	$I_{c2}$ (nA)  & 40.1 \\
	$I_{c3}$ (nA) & 31.1 \\
	$I_{c4}$ (nA) & 65.0 \\
	$I_{c5}$ (nA) &  65.2 \\
	$I_{c6}$ (nA) & 65.2 \\
	$I_{c7}$ (nA) &  65.0 \\
	$\omega_{J1}/2\pi$ (GHz) & 15.4 \\
   	$\omega_{J2}/2\pi$ (GHz)  & 19.9 \\
   	$\omega_{J3}/2\pi$ (GHz) & 15.4 \\
    	$\omega_{J4}/2\pi$ (GHz) & 32.3 \\
   	$\omega_{J5}/2\pi$ (GHz) & 32.4 \\
   	$\omega_{J6}/2\pi$ (GHz) & 32.4 \\
   	$\omega_{J7}/2\pi$ (GHz) & 32.3 \\
\end{tabular}
\end{ruledtabular}
\end{minipage}
\begin{minipage}{0.23\textwidth}
\begin{ruledtabular}
\begin{tabular}{ll}
	$r_{J8}=I_{c8}/\left[\left(I_{c4}+I_{c5}\right)/2\right]$ & \bf{0.3} \\
	$r_{J9}=I_{c9}/\left[\left(I_{c6}+I_{c7}\right)/2\right]$ & \bf{0.3} \\
	$I_{c8}$ (nA) & 19.5 \\
	$I_{c9}$ (nA) &  19.5 \\
	$\omega_{J8}/2\pi$ (GHz) & 9.69 \\
   	$\omega_{J9}/2\pi$ (GHz) & 9.69 \\
	$W_{11}/2\pi$ (MHz) & 221 \\
	$W_{22}/2\pi$ (MHz) & 204 \\
	$W_{33}/2\pi$ (MHz) & 221 \\
	$W_{44}/2\pi$ (MHz) & 219 \\
	$W_{55}/2\pi$ (MHz) & 218 \\
	$W_{66}/2\pi$ (MHz) & 218 \\
	$W_{77}/2\pi$ (MHz) & 219 \\
	$g_{12}/2\pi$ (MHz) & 2.34 \\
	$g_{13}/2\pi$ (MHz) & 0.13 \\
	$g_{14}/2\pi$ (MHz) & 283 \\
	$g_{15}/2\pi$ (MHz) & 6.93 \\
	$g_{16}/2\pi$ (MHz) & 1.12 \\
	$g_{17}/2\pi$ (MHz) & 0.28 \\
	$g_{23}/2\pi$ (MHz) & 2.35 \\
	$g_{24}/2\pi$ (MHz) & 7.14 \\
	$g_{25}/2\pi$ (MHz) & 284 \\
	$g_{26}/2\pi$ (MHz) & 284 \\
	$g_{27}/2\pi$ (MHz) & 7.14 \\
	$g_{34}/2\pi$ (MHz) & 0.28 \\
	$g_{35}/2\pi$ (MHz) & 1.12 \\
	$g_{36}/2\pi$ (MHz) & 6.94 \\
	$g_{37}/2\pi$ (MHz) & 283 \\
	$g_{45}/2\pi$ (MHz) & 43.2 \\
	$g_{46}/2\pi$ (MHz) & 3.02 \\
	$g_{47}/2\pi$ (MHz) & 0.54 \\
	$g_{56}/2\pi$ (MHz) & 38.8 \\
	$g_{57}/2\pi$ (MHz) & 3.02 \\
	$g_{67}/2\pi$ (MHz) & 43.2 \\
\end{tabular}
\end{ruledtabular}
\end{minipage}
\label{tab:paraDTC}
\end{table}
Parameter setting is shown in Table\ \ref{tab:paraDTC}. 
We choose bare transmon frequencies $\omega_i$ and capacitances $C_{ij}$ as design values.
By definition, $W_{ij}$ is uniquely determined by $C_{ij}$. 
The anharmonicity of the bare transmon $i$ is roughly given by $(-W_{ii})$\cite{Koch2007}. 
The coupling constant between the bare transmons $i$ and $j$, 
$g_{ij}$,  is proportinal to $W_{ij}$ (for $i\neq j$) as follows\cite{goto2022DTC, KG2023}: 
\begin{align}
	g_{ij}=\frac{W_{ij}}{2}\sqrt{\frac{(\omega_i + W_{ii})(\omega_j + W_{jj})}{W_{ii}W_{jj}}}.
\end{align}
The Josephson frequencies of transmons 1--7, $\omega_{Ji}$, are calculated as\cite{goto2022DTC, KG2023} 
\begin{align}
	\omega_{Ji} = \frac{\left(\omega_{i}+W_{ii}\right)^2}{8W_{ii}}. 
\end{align}
As for $\omega_{J8(9)}$, 
we set $r_{J8(9)}$, 
the ratio of $\omega_{J8(9)}$ to the average value of $\omega_{J4(6)}$ and $\omega_{J5(7)}$, 
to 0.3\cite{KG2023}. 
The critical current $I_{ci}$ is proportional to the Josephson frequency 
as 
$I_{ci}=\hbar\omega_{Ji}/\phi_0$. 

Here, we explain our parameter setting. 
(1) We set the detunings of the NN qubits, 
$(\rm{Q_1}, \rm{Q_2})$ and $(\rm{Q_2}, \rm{Q_3})$, 
to be larger than the absolute values of the qubit anharmonicities $W_{ii}$ $(i\in\{1,2,3\})$. 
Such a parameter regime, called the highly-detuned regime or out-of-straddling regime\cite{IBMSTC, goto2022DTC}, 
is preferable to suppress microwave crosstalk between the NN qubits, 
compared to the nearly resonant or in-the-straddling regime\cite{TheoriticalSTCcrosstalk2022}.
(2) 
From the perspective of suppressing microwave crosstalk, 
it is desirable to have a larger detuning between the NNN qubits $(\rm{Q_1}, \rm{Q_3})$ 
as well. 
However, 
alternating red-band (lower-frequency) and blue-band (higher-frequency) qubits
to keep the NN qubits highly detuned as shown in Fig.\ \ref{fig:redblue}, 
the NNN qubits belong to the same-frequency band and thus must nearly resonate.
Thus, in this work, 
we set the detuning between the NNN qubits to 
a sufficiently small value (10 MHz) compared to the anharmonicity. 
We will show that even with such a small detuning,
crosstalk between the NNN qubits is negligible in this system, but 
it is not the case for the STCs. 
The origin of this peformance difference will be discussed in Sec.\  \ref{sec:Discussion}.
(3)
In actual circuits, 
there are parasitic capacitances, 
e.g. $C_{12}$, $C_{13}$, and $C_{56}$, 
which are not shown in Fig.\ \ref{fig:DTC}. 
These capasitances may degrade the performance of the DTCs.
Therefore, they should not be ignored. 
Thus, 
we set all of the capacitances, including the parasitic ones, 
to correspond to the layout shown in Fig.\ \ref{fig:redblue}(b). 
We estimate them by the electromagnetic simulator ANSYS Q3D\cite{Ansys}. 
We will also discuss the case
where the NNN qubits are not on the horizontal line but on the diagonal line 
in Appendix \ref{appendix:diagonal}.
(4) 
We set 
$\omega_{i}$ ($i\in\{4,5,6,7\}$) and $r_{Ji}$ $(i\in\{8,9\})$
such that they lead to small residual $ZZ$ couplings 
and a high-performance adiabatic CZ gate operation. 

\section{Numerical results}
\label{sec:DTCthreequbits}
\begin{figure*}[t]
	\includegraphics[width=\textwidth]{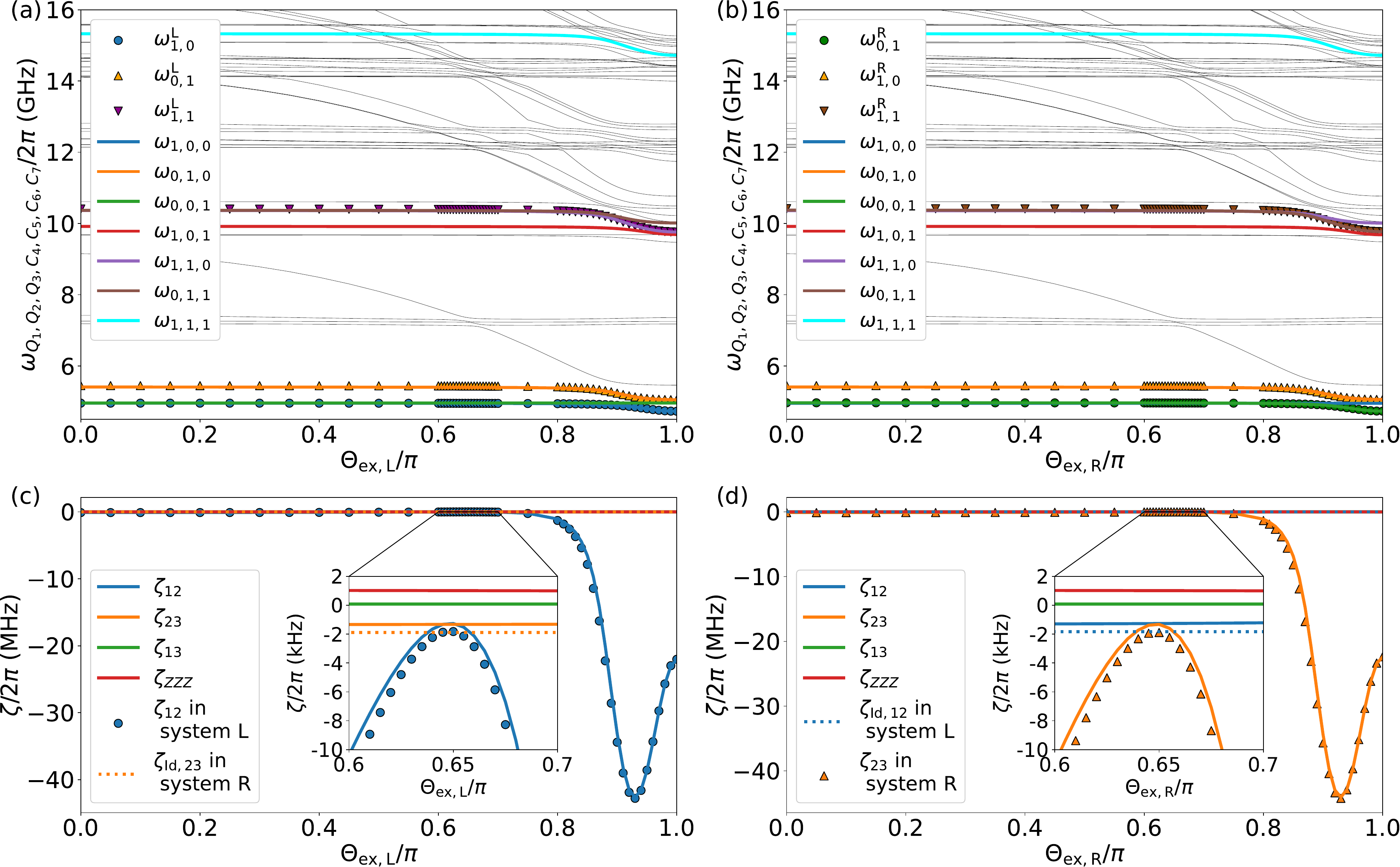}
	\caption{
	Energy levels and $ZZ$ coupling strengths of the systems in Fig.\ \ref{fig:DTC}. 
	(a) and (b) show
	eigenfrequencies $\omega_{Q_1,Q_2,Q_3,C_4,C_5,C_6,C_7}$ 
	as functions of $\Theta_{\rm ex, L}$ and $\Theta_{\rm ex, R}$, respectively. 
	The qubit frequencies of the three-qubit system, 
	$\omega_{Q_1,Q_2,Q_3}$ $(Q_i \in\{0,1\})$, 
	are highlighted by 
	being colored and bold,  
	and 
	the ones of the two-qubit system, 
	$\omega_{Q_i,Q_j}^{\mu}$ ($Q_i,Q_j\in\{ 0,1\}$ and $\mu\in\{{\rm L,R}\})$
	(see Appendix \ref{sec:IsolatedTwoQubitSubsystems}), 
	are shown by colored scatter plots. 
	Eigenfrequencies $\omega_{Q_1,Q_2,Q_3,C_4,C_5,C_6,C_7}$ other than 
	the above are represented by black thin curves.
	(c) and (d) represent
	$ZZ$ and $ZZZ$ couplings 
	as functions of $\Theta_{\rm ex, L}$ and $\Theta_{\rm ex, R}$, respectively.
	Solid curves represent the ones of the three-qubit system, 
	and scatter plots represent $\zeta_{12(23)}$ in the system L(R).
	The orange (blue) dotted horizontal line is $\zeta_{\rm Id,23(12)}$ in the system R(L). 
	Note that 
	$\Theta_{\rm ex, R}$ in the left column 
	and $\Theta_{\rm ex, L}$ in the right column 
	are fixed to $\Theta_{\rm Id, R}$ and $\Theta_{\rm Id, L}$, respectively.
	}
	\label{fig:DTCspectrumZZ}
\end{figure*}
\subsection{Idle point, $ZZ$ couplings, and $ZZZ$ coupling}
\label{sec:IdlingPoints}
\begin{table*}[t]
\caption{
\label{tab:gate}
Residual $ZZ$ couplings and $ZZZ$ coupling in the DTC architecture. 
}
\begin{ruledtabular}
\begin{tabular}{llllll}
	&  
	& $\zeta_{{\rm Id}, 12}/(2\pi)$ 
	& $\zeta_{{\rm Id}, 23}/(2\pi)$ 
	& $\zeta_{{\rm Id}, 13}/(2\pi)$ 
	& $\zeta_{{\rm Id}, ZZZ}/(2\pi)$ \\ 
	\hline
	& Three-qubit system 
	& -1 kHz 
	& -1 kHz 
	& 0 kHz 
	& 1 kHz \\
	\hline
	& Two-qubit subsystem 
	& -2 kHz 
	& -2 kHz 
	&  N/A
	&  N/A \\
\end{tabular}
\end{ruledtabular}
\label{tab:DTCZZ}
\end{table*}
\begin{table*}[t]
\caption{
\label{tab:gate}
Average gate fidelities of 30-ns CZ gates and 10-ns individual and simultaneous $\pi/2$ pulses. They are implemented by optimized pulses in the DTC architecture. 
}
\begin{ruledtabular}
\begin{tabular}{lllllllllllllll}
& \rm{Average gate fidelity (\%)}  
	& $\bar{F}_{CZ,12}$ 
	& $\bar{F}_{CZ,23}$ 
	& $\bar{F}_{\frac{\pi}{2},\{1\}}$ 
	& $\bar{F}_{\frac{\pi}{2},\{2\}}$ 
	& $\bar{F}_{\frac{\pi}{2},\{3\}}$
	& $\bar{F}_{\frac{\pi}{2},\{1,2\}}$
	& $\bar{F}_{\frac{\pi}{2},\{2,3\}}$
	& $\bar{F}_{\frac{\pi}{2},\{1,3\}}$
	& $\bar{F}_{\frac{\pi}{2},\{1,2,3\}}$ \\
	\hline
	& Three-qubit system
	& 100.00 & 100.00
	& 100.00 & 100.00 & 100.00 
	& 100.00 & 100.00 & 100.00
	& 100.00 \\
	\hline
	& Two-qubit system
	& 100.00 & 100.00
	& -------- & -------- & -------- 
	& -------- & -------- & -------- 
	& -------- \\ 
\end{tabular}
\end{ruledtabular}
\label{tab:DTCgate}
\end{table*}
Figure \ref{fig:DTCspectrumZZ}(a)[\ref{fig:DTCspectrumZZ}(b)]
shows $\omega_{Q_1,Q_2,Q_3,C_4,C_5,C_6,C_7}$ 
as functions of $\Theta_{\rm L(R)}$ 
when 
$\Theta_{\rm ex,R(L)}$ is fixed to $\Theta_{\rm Id,R(L)}\equiv0.65\pi\ (0.65\pi)$. 
Here, 
$\Theta_{\rm Id,R(L)}$
is an idle point of the isolated two-qubit subsystem L(R) 
where the $ZZ$ coupling 
between $\rm{Q_{1}}$ and $\rm{Q_{2}}$ ($\rm{Q_{2}}$ and $\rm{Q_{3}})$ is the closest to 0 
(see Appendix \ref{sec:IsolatedTwoQubitSubsystems} for the details). 
At the end of this subsection, 
we will explain the validity that 
the idle point of the three-qubit system 
can be chosen as 
$(\Theta_{\rm ex,L},\Theta_{\rm ex,R})=(\Theta_{\rm Id,L},\Theta_{\rm Id,R})$. 
The qubit frequencies of the three-qubit system, 
$\omega_{Q_1,Q_2,Q_3}$ $(Q_i \in\{0,1\})$, 
are highlighted by 
being colored and bold,  
and 
the ones of the two-qubit system, 
$\omega_{Q_i,Q_j}^{\mu}$ ($Q_i,Q_j\in\{ 0,1\}$ and $\mu\in\{{\rm L,R}\})$
(see Appendix \ref{sec:IsolatedTwoQubitSubsystems}), 
are shown by scatter plots. 
All of them are needed for calculations of $ZZ$ and $ZZZ$ couplings below. 
We find that 
$\omega_{1,0,0}$, $\omega_{0,1,0}$, $\omega_{0,0,1}$, 
$\omega_{1,1,0}$, and $\omega_{0,1,1}$
are  
in good agreement with the corresponding ones 
in the two-qubit systems.  
The black thin curves are frequencies of the states other than the computational state. 
They cannot be ignored to account for leakage errors during gate simulations. 

The $ZZ$ coupling between ${\rm Q}_i$ and ${\rm Q}_j$ 
is denoted by $\zeta_{ij}$ ($i<j$). 
The ones between the NN qubits, 
$\zeta_{12}$ and $\zeta_{23}$, 
are expressed as follows in the three-qubit system: 
\begin{align}
	\zeta_{12}
	&=\omega_{1,1,0} -\omega_{1,0,0}-\omega_{0,1,0}+\omega_{0,0,0}
	, \\
	\zeta_{23}
	&=\omega_{0,1,1} -\omega_{0,1,0}-\omega_{0,0,1}+\omega_{0,0,0}. 
\end{align}
We use the same notation 
, $\zeta_{12}$ and $\zeta_{23}$, 
for the corresponding $ZZ$ couplings in systems L and R, 
respectively (see Appendix \ref{sec:IsolatedTwoQubitSubsystems} for their definitions).
The scatter plot in Fig.\ \ref{fig:DTCspectrumZZ}(c)[\ref{fig:DTCspectrumZZ}(d)]
show $\zeta_{12(23)}$ in system L(R). 
The above $\Theta_{\rm Id,L(R)}$ has been 
identified as the point where $|\zeta_{12(23)}|$ takes the smallest value. 
We find that 
$\zeta_{12(23)}$ in the three-qubit system 
as a function of $\Theta_{\rm ex,L(R)}$ 
[the blue (orange) curve] 
is in good agreement with the one in system L(R). 
We also found that $\zeta_{23(12)}$ is almost independent of $\Theta_{\rm ex,L(R)}$ 
in the three-qubit system 
and its value is close to the one 
at the idle point in the isolated two-qubit system R(L) 
shown by the dotted horizontal line in Fig.\ \ref{fig:DTCspectrumZZ}(c)[\ref{fig:DTCspectrumZZ}(d)]. 
This independence is reasonable 
because $\Theta_{\rm ex,L(R)}$ 
is the parameter of DTC L(R) only connecting 
$\rm{Q_{1}}$ and $\rm{Q_{2}}$ ($\rm{Q_{2}}$ and $\rm{Q_{3}}$).

So far, we have considered $ZZ$ coupling between NN qubits, 
which also exists in two-qubit subsystems.
Here, 
we consider the 
$ZZ$ coupling between the NNN qubits $\zeta_{13}$ 
and the $ZZZ$ coupling, 
which exist only in $n(\geq 3)$-qubit systems. 
In the three-qubit system, 
they are expressed as follows: 
\begin{align}
	\zeta_{13}
	&=\omega_{1,0,1} -\omega_{1,0,0}-\omega_{0,0,1}+\omega_{0,0,0},\\
	\zeta_{ZZZ}
	&=\omega_{1,1,1} 
	-\left[\omega_{1,0,0}
	+\omega_{0,1,0}
	+\omega_{0,0,1}
	\right] \nonumber\\
	&
	\ \ \ \ \ \ \ 
	-\left[
	\zeta_{12}
	+\zeta_{13}
	+\zeta_{23}
	\right]
	+2\omega_{0,0,0}. 
\end{align}
From Figs.\ \ref{fig:DTCspectrumZZ}(c) and \ref{fig:DTCspectrumZZ}(d), 
we found that
$|\zeta_{13}|$ and $|\zeta_{ZZZ}|$
take negligibly small values of 
about 0--1 kHz in the wide range of the external flux. 

Thus, 
we conclude that 
$\zeta_{12}$ and $\zeta_{23}$ are almost unchanged from the ones of the isolated two-qubit systems
and also $\zeta_{13}$ and $\zeta_{ZZZ}$ are negligible.
Therefore, 
it is reasonable to take 
the idle point of the three-qubit system to 
$(\Theta_{\rm ex ,L}, \Theta_{\rm ex,R})=(\Theta_{\rm Id ,L}, \Theta_{\rm Id,R})$, 
as mentioned above. 
We represent 
$\zeta_{12}$, 
$\zeta_{23}$, 
$\zeta_{13}$, 
and $\zeta_{ZZZ}$
at the idle point simply 
as $\zeta_{{\rm Id}, 12}$, $\zeta_{{\rm Id}, 23}$, $\zeta_{{\rm Id, 13}}$, and $\zeta_{{\rm Id},ZZZ}$, respectively. 
Their values are summarized in Table \ref{tab:DTCZZ}.

\subsection{Gate performance}
\label{sec:gate}
To evaluate gate performance, 
we calculate an average gate fidelity $\bar{F}$. 
This is defined by averaging gate fidelities over uniformly distributed initial states
and calculated by 
the following formula\cite{fidelity,kueng2016comparing}:
\begin{align}
	\bar{F}
		=\frac{\left|{\rm tr}(\hat{U}_{\rm id}^{\dag}\hat{U}')\right|^2
		+{\rm tr}\left(\hat{U}'^{\dag}\hat{U}'\right)}{d(d+1)}, 
		\label{eq:fidelity}
\end{align}
where 
$d=2^{n}$ for an $n$-qubit system, 
$\hat{U}_{\rm id}$ is an ideal (target) gate operation matrix, 
and $\hat{U}'$ is an implemented gate operation matrix 
determined by numerical results
(see Appendices \ref{sec:Uprime} and \ref{sec:cariburation}).  

\subsubsection{Two-qubit gate: CZ gate}
\label{sec:CZ}
We aim to implement 30-ns CZ gates for qubit pairs 
$(\rm{Q_1}, \rm{Q_2})$ and $(\rm{Q_2}, \rm{Q_3})$. 
Their $\hat{U}_{\rm id}$ in the three-qubit system are given as follows:  
\begin{align}
	\hat{U}_{CZ,12}&=\hat{U}_{CZ}
		\otimes \hat{I}, \\ 
	\hat{U}_{CZ,23}&=
	\hat{I}\otimes \hat{U}_{CZ}, 
\end{align}
where $\hat{U}_{CZ}={\rm diag}(1,1,1,-1)$ and $\hat{I}$ is 
the $2\times2$ identity matrix. 
In two-qubit subsystems L and R, 
$\hat{U}_{CZ,12}$ and $\hat{U}_{CZ,23}$ are simply $\hat{U}_{CZ}$.
In the following, 
the average gate fidelities of $\hat{U}_{CZ,12}$ and $\hat{U}_{CZ,23}$ 
are denoted by $\bar{F}_{CZ,12}$ and $\bar{F}_{CZ,23}$, respectively.

To implement $\hat{U}_{CZ,12(23)}$ in the three-qubit system, 
we fix $\Theta_{\rm ex,R(L)}$ to $\Theta_{\rm Id,R(L)}$
and
apply a optimized flux pulse $\Theta_{\rm ex,L(R)}(t)$ 
(see Appendix \ref{sec:CZ_Optimization}).
We can successfully implement 30-ns CZ gates with average fidelities over 99.99\% 
for both the qubit pairs, 
as summarized in Table \ \ref{tab:DTCgate}. 
Comparing these fidelities with 
the corresponding ones in the isolated two-qubit subsystems, 
we find that 
the degradation of the CZ gate fidelity due to the increasing number of qubits 
from two to three 
is negligible in the 4-digit precision. 

\subsubsection{Single-qubit gate: $\pi/2$ pulse}
\label{sec:DTCsingle}
Next, we consider the single-qubit gate as well. 
Our targets are individual and simultaneous implementations of $10$-ns $\pi/2$ pulses. 
It leads to the following $\hat{U}_{\rm id}$: 
\begin{align}
	\hat{U}_{\pi/2,\{1\}}&= 
	\hat{U}_{\pi/2}
	\otimes \hat{I} \otimes \hat{I}, \\
	\hat{U}_{\pi/2,\{2\}}&= 
	\hat{I}\otimes
	\hat{U}_{\pi/2}
	\otimes \hat{I}, \\
	\hat{U}_{\pi/2,\{3\}}&= 
	\hat{I} \otimes \hat{I} \otimes
	\hat{U}_{\pi/2},  \\
	\hat{U}_{\pi/2,\{1,2\}}&= 
	\hat{U}_{\pi/2}
	\otimes \hat{U}_{\pi/2} \otimes \hat{I}, \\
	\hat{U}_{\pi/2,\{2,3\}}&= 
	\hat{I}
	\otimes \hat{U}_{\pi/2} \otimes \hat{U}_{\pi/2}, \\
	\hat{U}_{\pi/2,\{1,3\}}&= 
	\hat{U}_{\pi/2}
	\otimes \hat{I} \otimes \hat{U}_{\pi/2}, \\
	\hat{U}_{\pi/2,\{1,2,3\}}&= 
	\hat{U}_{\pi/2}
	\otimes \hat{U}_{\pi/2} \otimes \hat{U}_{\pi/2}, 
\end{align}
where 
\begin{align}
	\hat{U}_{\pi/2}=
	\frac{1}{\sqrt{2}}
	\begin{pmatrix}
 			  1 & -i  \\
 			  -i & 1  \\ 
	\end{pmatrix}.
\end{align}
The average gate fidelity of $\hat{U}_{\pi/2,k}$ is denoted by $\bar{F}_{\pi/2,k}$ 
for $k\in\{\{1\},\{2\},\{3\},\{1,2\},\{2,3\},\{1,3\},\{1,2,3\}\}$. 

To implement $\hat{U}_{\pi/2,k}$, 
we fix $\Theta_{\rm ex,\mu}$ to $\Theta_{\rm Id, \mu}$ 
and apply a driving voltage to ${\rm Q}_i$ for $i\in k$. 
This operation is described by the following Hamiltonian 
(see Appendix \ref{sec:derivation} for the derivation): 
\begin{align}
	\hat{H}_V &= 4\hbar\hat{\mathbf{n}}^{\rm T} W \hat{\mathbf{n}} 
	+ 
	\sum_{i \in k}\alpha_i \sum_{j=1}^7 \hbar W_{ij}\hat{n}_j 
	+\hat{U}(\Theta_{\rm Id,L}, \Theta_{\rm Id,R}), 
	\label{eq:Hamiltoniandrive}
\end{align}
where $\alpha_i$ is a dimensionless 
parameter controlled by 
$V_i$ (see Appendix \ref{sec:derivation}). 
Applying 
optimized $\alpha_i(t)$ (see Appendix \ref{sec:piover2_Optimization}), 
we can implement individual and simultaneous 10-ns $\pi/2$ pulses 
with average gate fidelities over 99.99\% as summarized in Table \ \ref{tab:DTCgate}.

\begin{figure*}[t]
\centering
\includegraphics[width=0.93\textwidth]{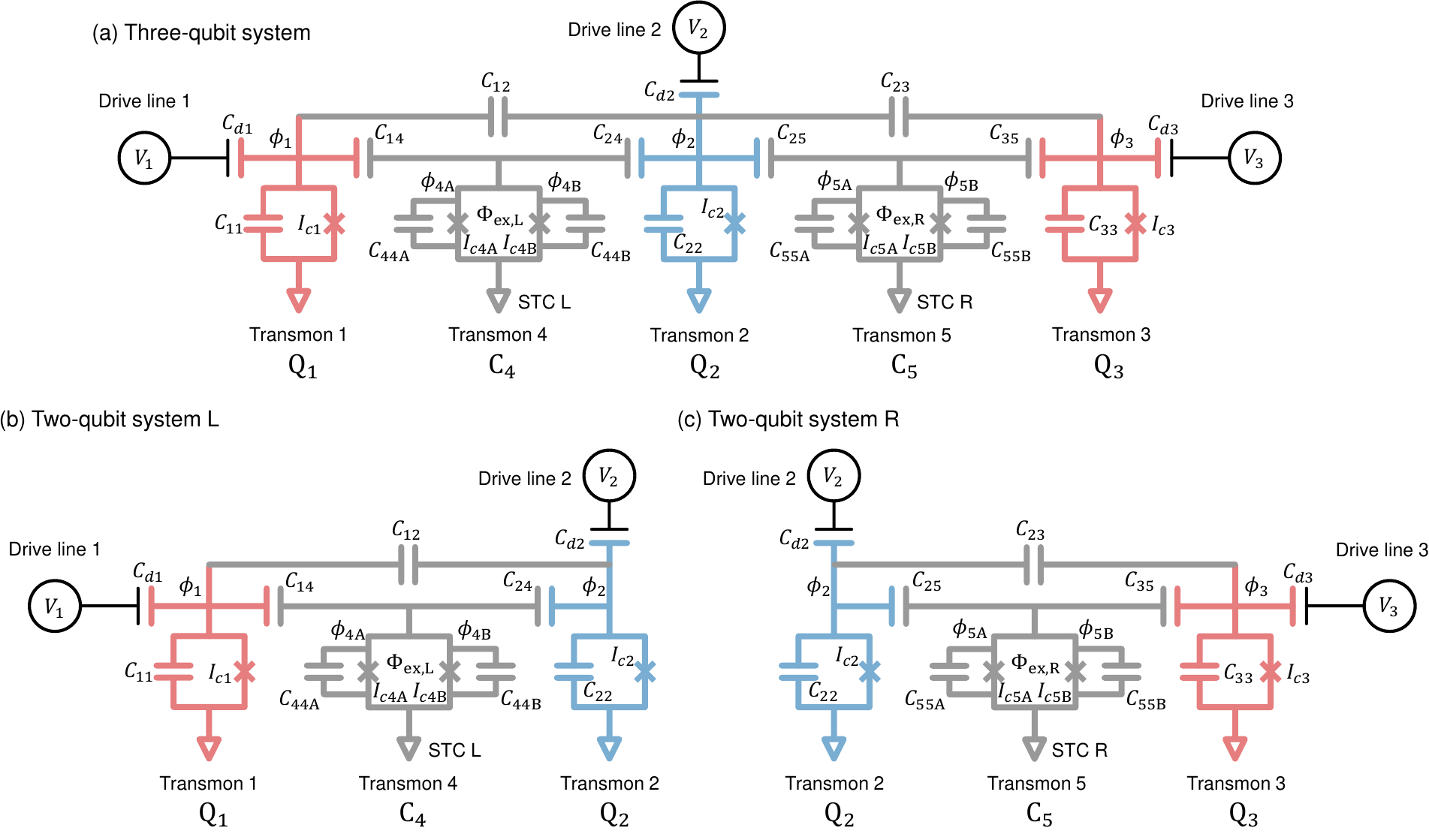}
\caption{
(a) Circuit diagram of the system with three qubits coupled via two STCs.  
(b) Circuit diagram of the two-qubit system L, 
which is the subsystem of (a).
(c) Similar diagram to (b) for the right side of (a). 
The colors of ${\rm Q}_i$ (red or blue) indicate 
which frequency bands (low or high) they belong to.  
}
\label{fig:STC}
\end{figure*}

In this section, 
we have shown that DTCs can not only reduce undesired residual couplings to several kHz 
but also enable implementations of fast CZ gates and individual and simultaneous $\pi/2$ pulses 
with average fidelities over 99.99\%, 
in the three-qubit system as well as in isolated two-qubit systems.
These results indicate that the DTC architecture is scalable 
for highly detuned fixed-frequency qubits.

\section{Comparison with STC architecture}
\label{sec:STC}
For comparison, here 
we consider the system 
where the DTCs in Fig.\ \ref{fig:DTC}(a) are replaced by STCs. 
We compare 
the DTC architecture with the STC one 
with respect to the following: 
\begin{enumerate}
	\item{Residual $ZZ$ and $ZZZ$ couplings.}
	\item{Average gate fidelities of optimized 30-ns CZ gates and of optimized 10-ns $\pi/2$ pulses. }
\end{enumerate}

\subsection{Circuit}
The circuit diagram is shown in Fig.\ \ref{fig:STC}(a). 
This system consists of five transmons. 
STC L (R) consists of a tunable-frequency transmon $\rm{C_4}$ ($\rm{C_5}$) 
and a capacitor $C_{12}$ ($C_{23}$). 
The tunable-frequency transmon ${\rm C}_i$ ($i\in\{4,5\}$) 
consists of 
two capacitors, 
$C_{ii{\rm A}}$ and $C_{ii{\rm B}}$, 
and a dc SQUID 
including two Josephson junctions 
with critical currents $I_{ci{\rm A}}$ and $I_{ci{\rm B}}$. 
Operator $\hat{\varphi}_{i\nu}$ and $\hbar\omega_{Ji\nu}=\phi_0 I_{ci\nu}$ 
are,
respectively, 
the phase difference operator 
and the Josephson energy corresponding to $I_{ci\nu}$ ($\nu\in\{{\rm A, B}\}$). 
In the following, 
we assume $C_{ii{\rm A}}=C_{ii{\rm B}}$
and use notations 
$C_{ii}=C_{ii{\rm A}}+C_{ii{\rm B}}$, 
$I_{ci}=I_{ci{\rm A}}+I_{ci{\rm B}}$, 
and $\omega_{Ji}=\omega_{Ji{\rm A}}+\omega_{Ji{\rm B}}$. 

\subsection{Hamiltonian of the superconducting circuit model}
We assume $V_i=0$ except for single-qubit gates. 
The Hamiltonian of this system is then written as follows 
(see Appendix \ref{sec:derivation} for the derivation): 
\begin{align}
	&\hat{H}
	= 4\hbar \mathbf{\hat{n}}^{\rm T} W \mathbf{\hat{n}}
	+ \left(
	\dot{\Theta}_{\rm ex,L}
	\mathbf{t}_{\rm SL}^{\rm T}
	+ \dot{\Theta}_{\rm ex,R}
	\mathbf{t}_{\rm SR}^{\rm T}
	\right)\hbar W\mathbf{\hat{n}}
	+\hat{U}, 
	\label{eq:HamiltonianSTC}
	\\
	&\hat{U} 
	=	- \sum_{i=1}^{3}\hbar\omega_{Ji}\cos(\hat{\varphi}_i) \nonumber\\
	&\ \ \ \ \ \ \ \ - \left[\hbar\omega_{J4{\rm A}}
	 +\hbar\omega_{J4{\rm B}}\cos(\Theta_{\rm ex,L})\right]
	 \cos(\hat{\varphi}_{4}) \nonumber\\
	 &\ \ \ \ \ \ \ \ -\hbar\omega_{J4{\rm B}}\sin(\Theta_{\rm ex,L})\sin(\hat{\varphi}_{4}) \nonumber\\
	 &\ \ \ \ \ \ \ \ - \left[\hbar\omega_{J5{\rm A}}
	 +\hbar\omega_{J5{\rm B}}\cos(\Theta_{\rm ex,R})\right]
	 \cos(\hat{\varphi}_{5}) \nonumber\\ 
	 &\ \ \ \ \ \ \ \ -\hbar\omega_{J5{\rm B}}\sin(\Theta_{\rm ex,R})\sin(\hat{\varphi}_{5}),
\end{align}
where 
$\hbar W = e^2M^{-1}/2$ 
with a capacitor matrix $M$ ($M_{ii}=\sum_{j=1}^5C_{ij}$ and $M_{ij}=-C_{ij}$ for $i\neq j$),  
\begin{align}
	\bm{t}_{\rm SL}^{\rm T}&=
	\frac{1}{2}
	\begin{pmatrix}
	  	-\frac{1}{\omega_{C_{14}}} & -\frac{1}{\omega_{C_{24}}} &
	  	 -\frac{1}{\omega_{C_{34}}} & \sum_{i=1}^5\frac{1}{\omega_{C_{i4}}} & 
	  	 -\frac{1}{\omega_{C_{45}}}\\
	\end{pmatrix},  
	\label{eq:tSL}\\
	\bm{t}_{\rm SR}^{\rm T}&=
	\frac{1}{2}
	\begin{pmatrix}
	  	-\frac{1}{\omega_{C_{15}}} & -\frac{1}{\omega_{C_{25}}} &
	  	 -\frac{1}{\omega_{C_{35}}} & -\frac{1}{\omega_{C_{45}}} & 
	  	 \sum_{i=1}^5\frac{1}{\omega_{C_{i5}}} \\
	\end{pmatrix},
	\label{eq:tSR}
\end{align}
and we have removed $\hat{\varphi}_{4{\rm B}(5{\rm B})}$ 
using the constraint $\phi_{4{\rm B}(5{\rm B})}=\phi_{4{\rm A}(5{\rm A})}-\Phi_{\rm ex,L(R)}$
and simply written $\hat{\varphi}_{4{\rm A}(5{\rm A})}$
as $\hat{\varphi}_{4(5)}$.
The matrix representation of the Hamiltonian can be obtained 
in the same manner as the DTC architecture.
We again choose the Cooper-pair number cutoff $N=10$ for sufficient convergence of energies.

The eigenfrequencies of the three-qubit system 
and corresponding eigenstates are denoted by
$\omega_{Q_1,Q_2,Q_3,C_4,C_5}$ and $\ket{Q_1,Q_2,Q_3,C_4,C_5}$ 
($Q_i, C_i \in {0,1,2,\cdots}$), respectively. 
Similarly to the DTCs, 
we also use notations 
of qubit frequencies
$\omega_{Q_1,Q_2,Q_3}\equiv\omega_{Q_1,Q_2,Q_3,0,0}$ 
and the corresponding eigenstates
$\ket{Q_1,Q_2,Q_3}\equiv\ket{Q_1,Q_2,Q_3,0,0}$. 

Note that almost all previous theoretical analyses of the STC reported in the literature 
have been based on the effective model
\cite{yan2018tunable,PRXSTC, RiggettiSTC,STCChinese, 
IBMSTC,YanSTCCZ,TheoriticalSTCcrosstalk2022}. 
Unlike them, we 
use the circuit model for higher accuracy, as well as the DTCs. 
To the best of our knowledge, 
this is the first estimation of $ZZ$ coupling between NNN qubits and $ZZZ$ coupling 
and also the first gate simulation based on the circuit model for systems with the STCs.

\subsection{Comparing conditions}
\begin{figure}[t]
\centering
\includegraphics[width=0.35\textwidth]{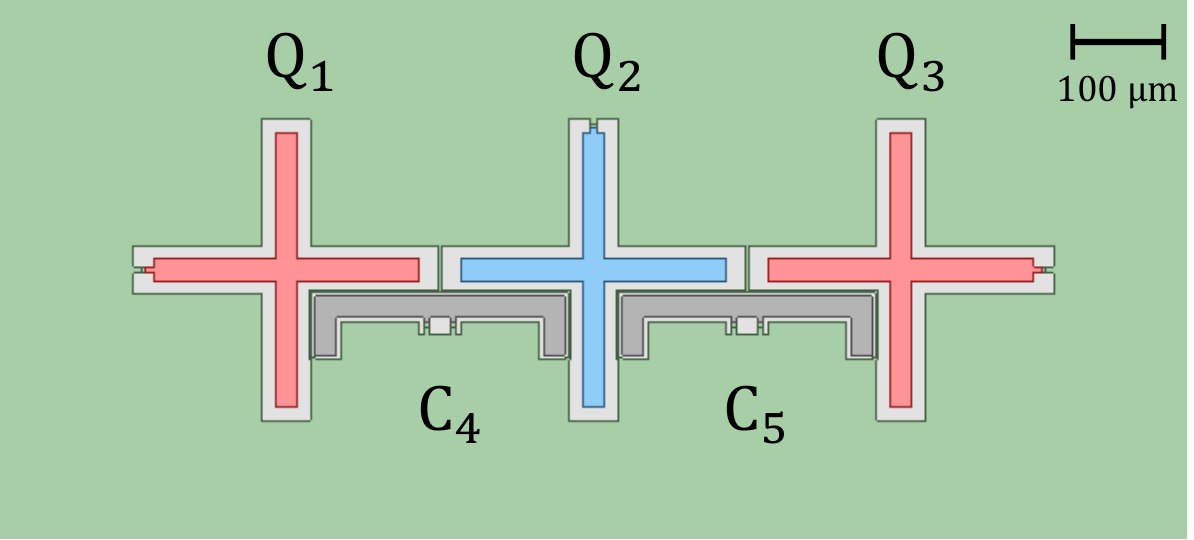}
\caption{
The scaled-up design of the circuit in Ref.\ \onlinecite{ETH}.
Its capacitances estimated by electromagnetic simulator ANSYS Q3D\cite{Ansys}
are summarized in Table \ref{tab:paraSTC}.
}
\label{fig:STClayout}
\end{figure}

The 
preferable parameter setting for the comparison between the STC and DTC architectures 
is non-trivial. 
This is because the STC and the DTC obey different operating principles, 
resulting in 
different parameter values desired for 
high performance.
Here, 
in order to 
make the comparison as fair as possible, 
we set the parameters as shown in Table \ref{tab:parametersSTC} , 
considering the following points.
(1) 
We set $\omega_i$ and $C_{ii}$ for $i\in\{1,2,3\}$ 
to be the same as in Table \ref{tab:paraDTC}, 
which means that qubit frequencies and anharmonicities are almost the same 
as in the DTC case.
(2) 
Under this condition, 
we set 
$C_{ij}$ that exists in the two-qubit subsystems  
and
$I_{c4{\rm B}(c5{\rm B})}/I_{c4\rm{A}(c5{\rm A})}$ 
to the almost same values as the corresponding ones in Ref.\ \onlinecite{ETH}, 
and we also set
$\omega_{4(5)}$ 
to the almost same value as the coupler idle frequency in Ref.\ \onlinecite{ETH}. 
Reference \onlinecite{ETH} experimentally implemented a high-performance 
CZ gate (with a gate time of 38 ns and a fidelity of 97.9\%), 
using a parameter setting close to the above condition (1) 
, where qubits frequencies are 5.038 and 5.400 GHz and both qubits capacitances are 77.8 fF. 
Moreover, 
they also showed that the magnitude of the residual $ZZ$ coupling 
can be suppressed to 60--80 kHz by an STC, 
which is known to be one of the most $ZZ$-coupling suppressed results 
for highly detuned qubits where a fast CZ gate can be implemented. 
We will show that our numerical simulation reproduces their 
results well. 
(3) 
As for the rest $C_{ij}$ that does not exist in the two-qubit subsystems (e.g. $C_{13}, C_{15}$, and $C_{45}$),  
we choose feasible values estimated 
by electromagnetic simulator ANSYS Q3D\cite{Ansys} 
assuming the scaled-up design of the circuit in Ref.\ \onlinecite{ETH} (see Fig.\ \ref{fig:STClayout}). 
\begin{table}[t]
\caption{
\label{tab:parametersSTC}
Parameter setting for the STC circuits in Fig.\ \ref{fig:STC}. 
Bold values are design values. 
The others are calculated using them.}
\begin{minipage}{0.23\textwidth}
\vspace{-0.75cm}
\begin{ruledtabular}
\begin{tabular}{ll}
	$\omega_{1}/(2\pi)$ (GHz) & {\bf 5.0}\\ 
    	$\omega_{2}/(2\pi)$ (GHz)  & {\bf 5.5}\\ 
   	$\omega_{3}/(2\pi)$ (GHz) & {\bf 5.01}\\ 
  	$\omega_{4}/(2\pi)$ (GHz) & {\bf 7.7}\\ 
  	$\omega_{5}/(2\pi)$ (GHz) & {\bf 7.7}\\ 
	$C_{ii}$ (fF) for $i\in\{1,2,3\}$& {\bf 80.0}\\ 
    	$C_{ii{\rm A}}$, $C_{ii{\rm B}}$ (fF) & {\bf 30.0} \\
	$C_{12},C_{23}$ (fF)  & {\bf 0.5}\\ 
	$C_{13}$ (fF) & {\bf 0.03} \\
 	$C_{14},C_{24},C_{25},C_{35}$ (fF) & {\bf 6.5}\\ 
 	$C_{15},C_{34}$ (fF) & {\bf 0.05} \\
	$C_{45}$ (fF) & {\bf 0.25} \\
	$I_{c1}$ (nA) & 30.7 \\
	$I_{c2}$ (nA)  & 39.3 \\
	$I_{c3}$ (nA) & 30.8 \\
	$\omega_{J1}/(2\pi)$ (GHz) & 15.2 \\
   	$\omega_{J2}/(2\pi)$ (GHz)  & 19.5 \\
   	$\omega_{J3}/(2\pi)$ (GHz) & 15.3 \\
\end{tabular}
\end{ruledtabular}
\end{minipage}
\begin{minipage}{0.23\textwidth}
\begin{ruledtabular}
\begin{tabular}{ll}
	$I_{ci{\rm B}}/I_{ci{\rm A}}$ & $\bf{1/1.71}$\\ 
	$I_{ci{\rm A}}$ (nA) & 37.7 \\
	$I_{ci{\rm B}}$ (nA) & 22.0 \\
	$\omega_{Ji{\rm A}}/(2\pi)$ (GHz) & 18.7 \\
   	$\omega_{Ji{\rm B}}/(2\pi)$ (GHz) & 10.9 \\
	$W_{11}/2\pi$ (MHz) & 224 \\
	$W_{22}/2\pi$ (MHz) & 209 \\
	$W_{33}/2\pi$ (MHz) & 224 \\
	$W_{44}/2\pi$ (MHz) & 268 \\
	$W_{55}/2\pi$ (MHz) & 268 \\
	$g_{12}/2\pi$ (MHz) & 33.0 \\
	$g_{13}/2\pi$ (MHz) & 1.62 \\
	$g_{14}/2\pi$ (MHz) & 265 \\
	$g_{15}/2\pi$ (MHz) & 6.13 \\
	$g_{23}/2\pi$ (MHz) & 33.0 \\
	$g_{24}/2\pi$ (MHz) & 268 \\
	$g_{25}/2\pi$ (MHz) & 268 \\
	$g_{34}/2\pi$ (MHz) & 6.14 \\
	$g_{35}/2\pi$ (MHz) & 265 \\
	$g_{45}/2\pi$ (MHz) & 39.2 \\
\end{tabular}
\end{ruledtabular}
\end{minipage}
\label{tab:paraSTC}
\end{table}

\subsection{Numerical results}
\begin{figure*}[t]
	\includegraphics[width=\textwidth]{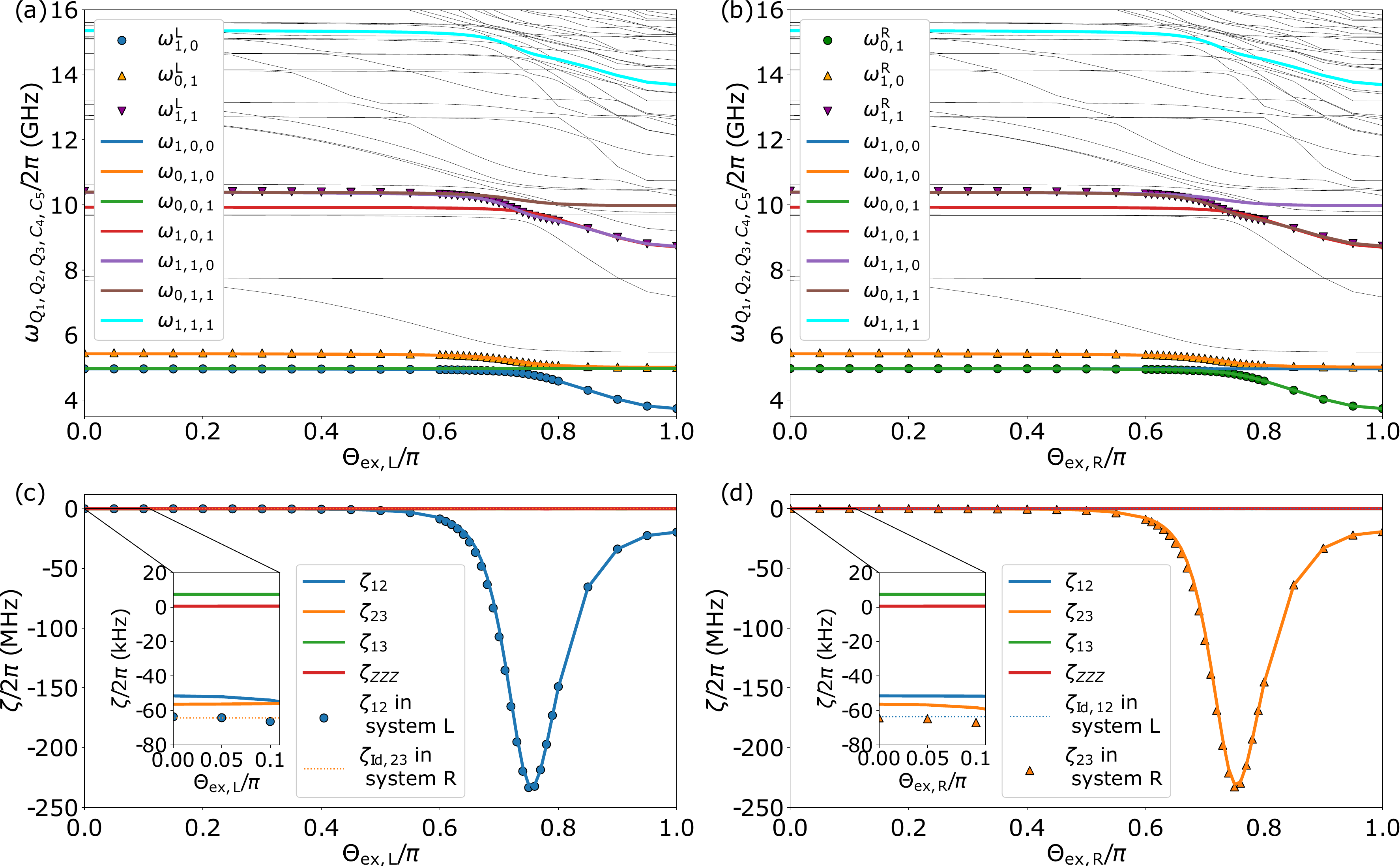}
	\caption{
	Energy levels and $ZZ$ coupling strengths of the systems in Fig.\ \ref{fig:STC}. 
	(a) and (b) show
	eigenfrequencies $\omega_{Q_1,Q_2,Q_3,C_4,C_5}$ 
	as functions of $\Theta_{\rm ex, L}$ and $\Theta_{\rm ex, R}$, respectively. 
	The qubit frequencies of the three-qubit system, 
	$\omega_{Q_1,Q_2,Q_3}$ $(Q_i \in\{0,1\})$, 
	are highlighted by 
	being colored and bold,  
	and 
	the ones of the two-qubit system, 
	$\omega_{Q_i,Q_j}^{\mu}$ ($Q_i,Q_j\in\{ 0,1\}$ and $\mu\in\{{\rm L,R}\})$
	(see Appendix \ref{sec:IsolatedTwoQubitSubsystems}), 
	are shown by colored scatter plots. 
	Eigenfrequencies $\omega_{Q_1,Q_2,Q_3,C_4,C_5}$ other than 
	the above are represented by black thin curves.
	(c) and (d) represent
	$ZZ$ and $ZZZ$ couplings 
	as functions of $\Theta_{\rm ex, L}$ and $\Theta_{\rm ex, R}$, respectively.
	Solid curves represent the ones of the three-qubit system, 
	and scatter plots represent $\zeta_{12(23)}$ in the system L(R).
	The Orange (blue) dotted horizontal line is $\zeta_{\rm Id,23(12)}$ in the system R(L). 
	Note that 
	$\Theta_{\rm ex, R}$ in the left column 
	and $\Theta_{\rm ex, L}$ in the right column 
	are fixed to $\Theta_{\rm Id, R}$ and $\Theta_{\rm Id, L}$, respectively.
	}
	\label{fig:STCspectrumZZ}
\end{figure*}
\begin{table*}[t]
\caption{
\label{tab:gate}
Residual $ZZ$ couplings and $ZZZ$ coupling in the STC architecture. 
}
\begin{ruledtabular}
\begin{tabular}{llllll}
	& 
	& $\zeta_{{\rm Id}, 12}/(2\pi)$ 
	& $\zeta_{{\rm Id}, 23}/(2\pi)$ 
	& $\zeta_{{\rm Id}, 13}/(2\pi)$ 
	& $\zeta_{{\rm Id}, ZZZ}/(2\pi)$ \\
	\hline
	& Three-qubit system 
	& -52 kHz 
	& -56 kHz 
	& 8 kHz 
	& 1 kHz \\ 
	\hline
	& Two-qubits system 
	& -64 kHz 
	& -64 kHz 
	&  N/A
	&  N/A \\ 
\end{tabular}
\end{ruledtabular}
\label{tab:STCZZ}
\end{table*}
\subsubsection{Idle point, $ZZ$ couplings, and $ZZZ$ coupling}
Figure \ref{fig:STCspectrumZZ}(a)[\ref{fig:STCspectrumZZ}(b)] 
shows $\omega_{Q_1,Q_2,Q_3,C_4,C_5}$ as functions of $\Theta_{\rm L(R)}$
when $\Theta_{\rm ex,R(L)}$ is fixed 
to the idle point of the isolated two-qubit subsystems L(R) 
$\Theta_{\rm Id,R(L)}\equiv0 (0)$
(see Appendix \ref{sec:IsolatedTwoQubitSubsystems}). 
The qubit frequencies of the three-qubit system, 
$\omega_{Q_1,Q_2,Q_3}$ $(Q_i \in\{0,1\})$, 
are highlighted by being colored and bold,  
and 
the ones of the two-qubit system, 
$\omega_{Q_i,Q_j}^{\mu}$ ($Q_i,Q_j\in\{ 0,1\}$ and $\mu\in\{{\rm L,R}\})$ 
are shown by scatter plots. 
All of them are needed for calculations of $ZZ$ and $ZZZ$ couplings below. 
Similarly to the DTC architecture, 
we find that the corresponding qubit frequencies of the three-qubit 
and two-qubit systems are in good agreement with each other.
The black thin curves represent the states other than the computational ones. 
They are necessary for accounting leakage errors during gate simulations.

Figure \ref{fig:STCspectrumZZ}(c)[\ref{fig:STCspectrumZZ}(d)] 
shows the $ZZ$ and $ZZZ$ couplings. 
Solid curves represent the ones of the three-qubit system, 
and scatter plots represent the ones of the two-qubit system. 
We find that $\zeta_{12}$ and $\zeta_{23}$, 
are almost unchanged from the ones of the isolated two-qubit systems 
and $\zeta_{13}$ and $\zeta_{ZZZ}$ are negligibly small (about 10 and 0 kHz, respectively). 
These results show that STCs can suppress the residual couplings even in three-qubit systems. 
Therefore, 
we can take the idle point of the three-qubit system to 
$(\Theta_{\rm ex ,L}, \Theta_{\rm ex,R})=(\Theta_{\rm Id ,L}, \Theta_{\rm Id,R})$. 
However, as in the two-qubit system,
their magnitudes shown in Table \ref{tab:STCZZ} 
are about an order of magnitude larger 
than the corresponding ones of the DTC architecture. 
These results show 
that the DTC is superior to 
the STC in terms of the residual couplings in three-qubit systems, 
as well as in two-qubit systems.
\begin{table*}[t]
\caption{
\label{tab:gate}
Average gate fidelities of 30-ns CZ gates and 10-ns individual and simultaneous $\pi/2$ pulses. They are implemented by optimized pulses in the STC architecture. 
}
\begin{ruledtabular}
\begin{tabular}{lllllllllllllll}
	& \rm{Average gate fidelity (\%)}  
	& $\bar{F}_{CZ,12}$ 
	& $\bar{F}_{CZ,23}$ 
	& $\bar{F}_{\frac{\pi}{2},\{1\}}$ 
	& $\bar{F}_{\frac{\pi}{2},\{2\}}$ 
	& $\bar{F}_{\frac{\pi}{2},\{3\}}$
	& $\bar{F}_{\frac{\pi}{2},\{1,2\}}$
	& $\bar{F}_{\frac{\pi}{2},\{2,3\}}$
	& $\bar{F}_{\frac{\pi}{2},\{1,3\}}$
	& $\bar{F}_{\frac{\pi}{2},\{1,2,3\}}$ \\
	\hline
	& Three-qubit system
	& 99.81 & 99.38 
	& 99.80 & 100.00 & 99.79
	& 99.80 & 99.79 & 99.63
	& 99.62 
	\\ 
	\hline
	& Two-qubit system 
	& 100.00  & 100.00 
	& --------  & --------  & -------- 
	& --------  & --------  & --------
	& --------  
	\\ 
\end{tabular}
\end{ruledtabular}
\label{tab:STCgate}
\end{table*}

\subsubsection{Gate performance}
\label{sec:STCgate}
Gate simulation results are summarized in Table \ \ref{tab:STCgate}. 
We find that 
30-ns CZ gates with average fidelities of 99.99\%
cannot be achieved in the three-qubit system. 
This is in contrast to the corresponding fidelities of the two-qubit subsystems, 
which exceeds 99.99\%.  
The fidelity degradation of the CZ gate due to 
the scale-up from two qubits to three
is about 0.2\%--0.6\%, 
which is roughly 10 times larger than the corresponding one in the DTC architecture. 
We also find that
10-ns $\pi/2$-pulses with average fidelities over 99.99\% 
cannot be implemented except for the $\rm{Q_2}$. 
The infidelities of the $\pi/2$ pulses for $\rm{Q_1}$ and $\rm{Q_3}$, 
are about 0.2\%.
This is also roughly 10 times larger than the corresponding ones of the system of the DTC.   

From these results, 
we conclude that
the DTC 
exhibits higher scalability than the STC, 
as the tunable coupler for highly detuned fixed-frequency qubits.

\section{Discussion}
\label{sec:Discussion}
As demonstrated in the above two sections, 
the DTCs can implement more accurate gate operations 
in the three-qubit system 
than the STCs.
This difference 
mainly comes from the larger 
parasitic coupling of the NNN qubits $g_{13}$ in the STC architecture 
than that in the DTC architecture 
(see Tables \ref{tab:paraDTC} and \ref{tab:paraSTC}). 
Since 
tunable couplers can, in principle, only
cancel couplings 
between NN qubits coupled via them, 
and have
no mechanism to cancel the other couplings, 
$g_{13}$ is always active. 
It means that 
a larger $|g_{13}|$ will lead to 
more serious errors in the $\rm{Q}_1$-$\rm{Q}_3$ subspace, 
such as microwave crosstalk. 
In fact, these errors are more pronounced for the STC than for the DTC 
(see Appendix \ref{appendix:errors}).

The above difference in $g_{13}/(2\pi)$ (0.13 MHz and 1.62 MHz for the DTC and the STC architecture, respectively) 
arises from the structure of the coupler and the coupling cancellation mechanism. 
Since the STCs utilize the relatively large direct capacitances between the NN qubits, 
$C_{12}$ and $C_{23}$, 
for canceling residual couplings\cite{yan2018tunable}, 
NN qubits tend to be close to each other.
This makes 
NNN qubits close as well, 
and thus the capacitance between the NNN qubits $C_{13}$ tends 
to be large (0.03 fF). 
On the other hand, 
the DTC does not require $C_{12}$ and $C_{23}$, 
allowing for greater distances 
between NN qubits and hence between NNN qubits, 
leading to a smaller $C_{13}$ (0.003 fF). 
This difference in $C_{13}$ is a major reason 
for the difference in $g_{13}$.

Note that even if $C_{13}$ is zero, 
there is still an effective capacitance, 
and hence the $g_{13}$ is not zero:  
0.04 MHz and 0.71 MHz for the DTC and the STC architecture, respectively. 
The one in the STC still remains about a half of the original value. 

In an ideal limit where all parasitic capacitances are zero, 
$g_{13}/(2\pi)$ of the DTC and STC 
become 0.00 MHz and 0.38 MHz, respectively. 
This means that in the DTC architecture, 
$g_{13}$ arises through 
parasitic capacitances 
and vanishes in the ideal situation. 
On the other hand, even in this ideal situation, 
$g_{13}$ in the STC architecture is about a half of that before taking the limit.

The remaining $g_{13}$ in the STC architecture is determined by two factors. 
One of them is the direct couplings between NN qubits, namely $C_{12}$ and $C_{23}$.  
In the STC architecture, 
they are large for canceling residual couplings as mentioned above, 
and thus this contribution does not vanish even in
the above ideal situation with no parasitic capacitances. 
On the other hand, 
the DTC does not rely on $C_{12}$ and $C_{13}$, that is, 
$C_{12}$ and $C_{23}$ are parasitic capacitances for the DTC architecture, 
and hence the contribution is negligible. 

The other factor is the indirect couplings through a coupler transmon. 
This contribution can be estimated by considering the case where $C_{12}$ and $C_{23}$ are zero, 
resulting in 0.11 MHz for the STC architecture. 
In the DTC architecture, 
corresponding contribution is negligible because the coupling between the two transmons inside DTC L(R) is nearly disconnected due to the small parasitic capacitance $C_{45}$($C_{67}$). 

In summary, 
the higher performance of the DTC architecture than the STC one 
mainly comes from 
the fact that 
the DTC has 
the advantage that it can reduce the NNN coupling, $g_{13}$, 
compared to the STC from various points.

\section{Summary}
\label{sec:summary}
In this paper, 
we have analyzed the system of three fixed-frequency-transmon qubits coupled via two DTCs, 
by numerical simulation using the circuit model, 
where the NN qubits are highly detuned and NNN qubits are nearly resonant. 
As a result, 
we have found that 
the DTC can suppress the residual $ZZ$ couplings and $ZZZ$ coupling, 
as well as in isolated two-qubit systems. 
Moreover, 
we have succeeded in implementing 30-ns CZ gates 
and not only individual but also simultaneous 10-ns $\pi/2$ pulses with average fidelities over 99.99\%. 
The degradation of the CZ gate fidelities due to the increasing number of qubits 
from two to three is 
negligible in the 4-digit precision. 

For comparison, 
additional simulations of 
the system where DTCs were replaced by STCs 
have been performed. 
Then, 
we have found 
that 
both suppressing residual couplings to several kHz 
and achieving two-qubit and single-qubit gates with fidelities of 99.99\% 
are highly challenging for the STCs. 
Due to the structure of STCs, 
the NNN-qubit coupling strength $g_{13}$ tends to be larger than DTCs, 
and hence causes more severe errors.

From these results, 
we have concluded that the DTC architecture is more promising 
than the STC architecture 
for realizing high-performance, scalable superconducting quantum computers. 
We expect that our results will be confirmed experimentally in the near future.





\appendix{}
\section{Derivation of the Hamiltonian}
\label{sec:derivation}

\subsection{DTC architecture}
The Lagrangian $L$ describing the system in Fig.\ \ref{fig:DTC}(a) is given as follows: 
\begin{align}
	&L=K-U, \\
	&K={\sum_{i=1}^7\frac{C_{ii}}{2}\dot{\phi}_i^2} 
	+{\sum_{i=1}^3\sum_{j=i+1}^7\frac{C_{ij}}{2}(\dot{\phi}_i-\dot{\phi}_j)^2}
	+\sum_{i=4}^5\sum_{j=6}^7\frac{C_{ij}}{2}(\dot{\phi}_i-\dot{\phi}_j)^2
	\nonumber\\
	&\ \ \ \ 
	+\frac{C_{45}}{2}\dot{\phi}_8^2
	+\frac{C_{67}}{2}\dot{\phi}_9^2 +\sum_{i=1}^3\frac{C_{di}}{2}(\dot{\phi}_i-V_i)^2, \\
	&U = -\sum_{i=1}^7 \hbar\omega_{J_i}\cos\varphi_i 
		-\hbar\omega_{J8}\cos(\varphi_{8})  
		-\hbar\omega_{J9}\cos(\varphi_{9}). 
\end{align}
Using the constraints 
$\phi_{8}=\phi_{5}-\phi_{4}-\Phi_{\rm ex,L}$ and $\phi_{9}=\phi_{7}-\phi_{6}-\Phi_{\rm ex,R}$
and neglecting the constant terms, 
the $K$ and $U$ are rewritten as follows: 
\begin{align}
	&K=\frac{1}{2}\dot{\mathbf{\phi}}^{\rm T}M\dot{\mathbf{\phi}}
	-\mathbf{q}_D^{\rm T}\dot{\mathbf{\phi}}
	-\frac{e}{4}\mathbf{\alpha}^{\rm T}\dot{\mathbf{\phi}}, 
	\label{eq:kineticDTC}
	\\
	&U = -\sum_{i=1}^7 \hbar\omega_{J_i}\cos\varphi_i \nonumber\\
		&\ \ \ \ \ \ \ \ 
		-\hbar\omega_{J8}\cos(\varphi_{5}-\varphi_4-\Theta_{\rm ex,L})  \nonumber\\
		&\ \ \ \ \ \ \ \ 
		-\hbar\omega_{J9}\cos(\varphi_{7}-\varphi_6-\Theta_{\rm ex,R}), 
\end{align}
where 
\begin{align}
\mathbf{\phi}^{\rm T}&=(\phi_1, \phi_2, \phi_3, \phi_4, \phi_5, \phi_6, \phi_7), \\
\mathbf{q}_D^{\rm T}
&=(0, 0, 0, -C_{45}\dot{\Phi}_{\rm ex,L}, C_{45}\dot{\Phi}_{\rm ex,L}, 
-C_{67}\dot{\Phi}_{\rm ex,R}, C_{67}\dot{\Phi}_{\rm ex,R}) \nonumber\\
&=\frac{e}{4}(\dot{\Theta}_{\rm ex,L}\mathbf{t}_{\rm DL}^{\rm T}+\dot{\Theta}_{\rm ex,R}\mathbf{t}_{{\rm DR}}^{\rm T}), \\
\mathbf{\alpha}^{\rm T}&=(\alpha_1, \alpha_2, \alpha_3, 0, 0, 0, 0), 
\end{align} 
with
$\alpha_i = 2eV_{i}/(\hbar\omega_{C_{{\rm d}i}})$ for $i\in\{1,2,3\}$ 
and  
$\mathbf{t}_{\rm D\mu}^{\rm T}$ for $\mu \in\{\rm L, R\}$ defined 
by Eqs.\ (\ref{eq:tDL}) and (\ref{eq:tDR}).

The Hamiltonian is obtained by the Legendre transformation of $L$ as
\begin{align}
	H&=\mathbf{Q}^{\rm T}\dot{\mathbf{\phi}}-L \nonumber\\
	&=\frac{1}{2}\mathbf{Q}^{\rm T}M^{-1}\mathbf{Q}+\mathbf{q}_D^{\rm T}M^{-1}\mathbf{Q}
	+\frac{e}{4}\mathbf{\alpha}^{\rm T} M^{-1} \mathbf{Q}
	+U
	, \\
	\mathbf{Q}&=
	\frac{\partial L}{\partial \dot{\mathbf{\phi}}}=M\dot{\mathbf{\phi}}-\mathbf{q}_D
	-\frac{e}{4}\mathbf{\alpha}, 
\end{align}
where $\mathbf{Q}$ denotes charge variables 
that are canonical conjugate variables for the flux variables $\mathbf{\phi}$. 
Introducing the Cooper-pair number variables as $\mathbf{n}=\mathbf{Q}/(2e)$, 
$H$ is rewritten as follows: 
\begin{align}
	H &= 4\hbar\mathbf{n}^{\rm T} W \mathbf{n}
	+ 
	\left(
	\dot{\Theta}_{\rm ex,L}
	\mathbf{t}_{\rm DL}^{\rm T}
	+ \dot{\Theta}_{\rm ex,R}
	\mathbf{t}_{\rm DR}^{\rm T}\right)
	\hbar W\mathbf{n}
	+\mathbf{\alpha}^{\rm T} \hbar W \mathbf{n} 
	+ U.
\end{align}
By canonical quantization procedure, 
namely, 
replacing the classical values $\mathbf{n}$ and $\mathbf{\phi}$
, respectively, with operators $\hat{\mathbf{n}}$ and $\hat{\mathbf{\phi}}$ 
satisfying the canonical commutation relation $[\hat{\varphi}_i, \hat{n}_j]=i\delta_{i,j}$, 
we obtain the quantized Hamiltonian: 
\begin{align}
	\hat{H} &= 4\hbar\hat{\mathbf{n}}^{\rm T} W \hat{\mathbf{n}}
	+ 
	\left(
	\dot{\Theta}_{\rm ex,L}
	\mathbf{t}_{\rm DL}^{\rm T}
	+ \dot{\Theta}_{\rm ex,R}
	\mathbf{t}_{\rm DR}^{\rm T}\right)
	\hbar W\hat{\mathbf{n}} \nonumber\\
	& \ \ \ \ +\mathbf{\alpha}^{\rm T} \hbar W \hat{\mathbf{n}} 
	+ \hat{U}.
\end{align}
As special cases, 
Eq.\ (\ref{eq:Hamiltonian}) is obtained when $\bm{\alpha}=\bm{0}$ 
and Eq.\ (\ref{eq:Hamiltoniandrive}) is obtained 
when $(\Theta_{\rm ex,L}, \Theta_{\rm ex,R})=(\Theta_{\rm Id,L}, \Theta_{\rm Id,R})$, 
$\alpha_i\neq0$, and $\alpha_j=0$ for $j\neq i$. 
The Hamiltonian of the isolated two qubit systems $\hat{H}^{\rm L(R)}$ can also be derived in a similar manner.

\subsection{STC architecture}
The Lagrangian $L$ describing the system in Fig.\ \ref{fig:STC}(a) 
is given as follows: 
\begin{align}
	&L=K-U, \\
	&K=\sum_{i=1}^3\frac{C_{ii}}{2}\dot{\phi}_i^2
	+\sum_{i=4}^5 \sum_{\nu\in\{\rm A, B\}}
	\frac{C_{ii\nu}}{2}\dot{\phi}_{i\nu}^2 
	+\sum_{i=1}^{2}\sum_{j=i+1}^3\frac{C_{ij}}{2}(\dot{\phi}_j-\dot{\phi}_i)^2 \nonumber\\
	&\ \ \ \ \ \ \ +\sum_{i=1}^{3}\sum_{j=4}^5\sum_{\nu\in\{\rm A,B\}}
	\frac{1}{2}\frac{C_{ij}}{2}(\dot{\phi}_{j\nu}-\dot{\phi}_i)^2 \nonumber\\
	&\ \ \ \ \ \ \ +\sum_{\nu_1,\nu_2\in\{\rm A, B\}}
	\frac{1}{2}\frac{C_{45}}{4}(\dot{\phi}_{5\nu_1}-\dot{\phi}_{4\nu_2})^2 
	+\sum_{i=1}^3\frac{C_{di}}{2}(\dot{\phi}_i-V_i)^2, \\
	&U=-\sum_{i=1}^3\hbar\omega_{Ji}\cos{\varphi_i}
	-\sum_{i=4}^5\sum_{\nu\in\{\rm{A,B}\}}\hbar\omega_{Ji\nu}\cos\varphi_{i\nu},
\end{align}
Using the constraints 
$\phi_{4{\rm B}}=\phi_{4{\rm A}}-\Phi_{\rm ex,L}$ and $\phi_{5{\rm B}}=\phi_{5{\rm A}}-\Phi_{\rm ex,R}$, 
the addition theorem, 
and neglecting the constant terms, 
$K$ and $U$ are rewritten as follows: 
\begin{align}
	&K=\frac{1}{2}\dot{\mathbf{\phi}}^{\rm T}M\dot{\mathbf{\phi}}
	-\mathbf{q}_S^{\rm T}\dot{\mathbf{\phi}} 
	-\frac{e}{4}\mathbf{\alpha}^{\rm T}\dot{\mathbf{\phi}},
	\\
	U
	&=	- \sum_{i=1}^{3}\hbar\omega_{Ji}\cos(\varphi_i) \nonumber\\
	&\ \ \ \ \ - \left[\hbar\omega_{J4{\rm A}}
	 +\hbar\omega_{J4{\rm B}}\cos(\Theta_{\rm ex,L})\right]
	 \cos(\varphi_{4}) \nonumber\\
	 &\ \ \ \ \ -\hbar\omega_{J4{\rm B}}\sin(\Theta_{\rm ex,L})\sin(\varphi_{4}) \nonumber\\
	 &\ \ \ \ \ - \left[\hbar\omega_{J5{\rm A}}
	 +\hbar\omega_{J5{\rm B}}\cos(\Theta_{\rm ex,R})\right]
	 \cos(\varphi_{5}) \nonumber\\ 
	 &\ \ \ \ \  -\hbar\omega_{J5{\rm B}}\sin(\Theta_{\rm ex,R})\sin(\varphi_{5}),
\end{align}
where 
\begin{align}
\mathbf{\phi}^{\rm T}&=(\phi_1, \phi_2, \phi_3, \phi_4, \phi_5), \\
\mathbf{q}_{S}^{\rm T}&=\frac{e}{4}(\dot{\Theta}_{\rm ex,L}\bm{t}_{{\rm SL}}^{\rm T}+, \dot{\Theta}_{\rm ex,R}\bm{t}_{{\rm SR}}^{\rm T}), \\
\mathbf{\alpha}^{\rm T}&=(\alpha_1, \alpha_2, \alpha_3, 0, 0), 
\end{align}
with $\mathbf{t}_{\rm S\mu}^{\rm T}$ for $\mu \in\{\rm L, R\}$ defined 
by Eqs.\ (\ref{eq:tSL}) and (\ref{eq:tSR}), 
and we have simply written $\phi_{4{\rm A}(5{\rm A})}$ as $\phi_{4(5)}$. 
We can obtain the Hamiltonian by the Legendre transformation of $L$ as 
\begin{align}
	H &= 4\hbar\mathbf{n}^{\rm T} W \mathbf{n}
	+ 
	\left(
	\dot{\Theta}_{\rm ex,L}
	\mathbf{t}_{\rm SL}^{\rm T}
	+ \dot{\Theta}_{\rm ex,R}
	\mathbf{t}_{\rm SR}^{\rm T}\right)
	\hbar W\mathbf{n}
	+\mathbf{\alpha}^{\rm T} \hbar W \mathbf{n} 
	+ U.
\end{align}
By canonical quantization of it, 
we obtain the following quantized Hamiltonian: 
\begin{align}
	\hat{H} &= 4\hbar\hat{\mathbf{n}}^{\rm T} W \hat{\mathbf{n}}
	+ 
	\left(
	\dot{\Theta}_{\rm ex,L}
	\mathbf{t}_{\rm SL}^{\rm T}
	+ \dot{\Theta}_{\rm ex,R}
	\mathbf{t}_{\rm SR}^{\rm T}\right)
	\hbar W\hat{\mathbf{n}} \nonumber\\
	& \ \ \ \ +\mathbf{\alpha}^{\rm T} \hbar W \hat{\mathbf{n}} 
	+ \hat{U}.
\end{align}
Equation (\ref{eq:HamiltonianSTC}) is obtained when $\mathbf{\alpha}=\bm{0}$.  
The Hamiltonian of the isolated two qubit systems $\hat{H}^{\rm L(R)}$ can also be derived in a similar manner.

\section{Matrix representation of the Hamiltonian}
\label{sec:MatrixRepresentation}
The matrix representations of the operators 
$\hat{n}_i$, $\cos\hat{\varphi}_i$, and $\sin\hat{\varphi}_i$
in the basis of $\hat{n}_i$ eigenfunctions are given as follows: 
\begin{align}
	\hat{n}_i&=
		\begin{pmatrix}
			  -N & \ & \ \\
 			  \ &\ddots &\ \\ 
 			  \ &\ &N \\
		\end{pmatrix}, \\
	\cos\hat{\varphi}_i&=\frac{1}{2}
		\begin{pmatrix}
			  \ &1 &\ &\ \\
 			  1 &\ &\ddots &\ \\ 
 			  \ &\ddots &\ &1 \\
 			  \ &\ &1 &\  \\
		\end{pmatrix}, \\
	\sin\hat{\varphi}_i&=\frac{1}{2i}
		\begin{pmatrix}
			  \ &-1 &\ &\ \\
 			  1 &\ &\ddots &\ \\ 
 			  \ &\ddots &\ &-1 \\
 			  \ &\ &1 &\  \\
		\end{pmatrix}, 
\end{align}
where $N$ is a cutoff for the Cooper-pair number. 
Each operator is expressed as a $(2N+1)\times(2N+1)$ matrix. 

Since the total system shown in Fig.\ \ref{fig:DTC}(a) is composed of seven subsystems (transmons), 
each operator in Eq.\ (\ref{eq:Hamiltonian}) is replaced by a tensor product of 
seven operators, such as 
$\hat{n}_1\otimes\hat{I}_2\otimes\hat{I}_3\otimes\hat{I}_4\otimes\hat{I}_5\otimes\hat{I}_6\otimes\hat{I}_7$, where $\hat{I}_i$ is the identity operator for the $i$th subsystem of 
$\hat{\varphi}_i$ and $\hat{n}_i$.
From the addition theorem, 
$\cos(\hat{\varphi}_{5}-\hat{\varphi}_4-\Theta_{\rm ex,L})$
and 
$\cos(\hat{\varphi}_{7}-\hat{\varphi}_6-\Theta_{\rm ex,R})$
are, respectively, expressed as follows:
\begin{align}
&\cos(\hat{\varphi}_{5}-\hat{\varphi}_4-\Theta_{\rm ex,L})\nonumber\\
&=\hat{I}_1\otimes\hat{I}_2\otimes\hat{I}_3 \nonumber\\
&\otimes
\Bigg\{
\cos\Theta_{\rm ex,L}
	\left[
		\cos(\hat{\varphi}_4)\otimes\cos(\hat{\varphi}_5)
		+\sin(\hat{\varphi}_4)\otimes\sin(\hat{\varphi}_5)
	\right] \nonumber\\
&+
\sin\Theta_{\rm ex,L}
	\left[
		\cos(\hat{\varphi}_4)\otimes\sin(\hat{\varphi}_5)
		-\sin(\hat{\varphi}_4)\otimes\cos(\hat{\varphi}_5)
	\right]
\Bigg\}\nonumber\\
&\otimes\hat{I}_6\otimes\hat{I}_7, 
\end{align}
\begin{align}
&\cos(\hat{\varphi}_{7}-\hat{\varphi}_6-\Theta_{\rm ex,R})\nonumber\\
&=\hat{I}_1\otimes\hat{I}_2\otimes\hat{I}_3\otimes\hat{I}_4\otimes\hat{I}_5 \nonumber\\
&\otimes
\Bigg\{
\cos\Theta_{\rm ex,R}
	\left[
		\cos(\hat{\varphi}_6)\otimes\cos(\hat{\varphi}_7)
		+\sin(\hat{\varphi}_6)\otimes\sin(\hat{\varphi}_7)
	\right] \nonumber\\
&+
\sin\Theta_{\rm ex,R}
	\left[
		\cos(\hat{\varphi}_6)\otimes\sin(\hat{\varphi}_7)
		-\sin(\hat{\varphi}_6)\otimes\cos(\hat{\varphi}_7)
	\right]
\Bigg\}.
\end{align}
In the matrix representation, 
$\hat{I}_i$ is given by the $(2N+1)\times(2N+1)$ identity matrix 
and the tensor product $\otimes$ is replaced by the Kronecker product of matrices.
Thus, we obtain a $(2N+1)^7\times (2N + 1)^7$ matrix representation of the Hamiltonian in Eq.\ (\ref{eq:Hamiltonian}). 

Similarly,  
we can obtain a $(2N+1)^5\times (2N + 1)^5$ matrix representation of the Hamiltonian in Eq.\ (\ref{eq:HamiltonianSTC}). 

\section{Dimension reduction technique}
\label{sec:simulation}
In this section, we introduce 
our numerical method to reduce the computational cost. 
Here we only explain 
the case of the three-qubit system with 
two DTCs. 
Other systems can also be treated similarly.

\subsection{Diagonalization}
The Hamiltonian in Eq.\ (\ref{eq:Hamiltonian}) 
can be divided as follows.
\begin{align}
	&\hat{H}=\hat{H}_{123}+\hat{H}_{45}+\hat{H}_{67}+\hat{H}_{\rm int}, \\
	&\hat{H}_{123}=\sum_{i=1}^3 
	\left(4\hbar W_{ii} \hat{n}_i^2-\hbar\omega_{J_i}\cos\hat{\varphi}_i\right), \\
	&\hat{H}_{45}=\sum_{i=4}^5 
	\left(4\hbar W_{ii} \hat{n}_i^2-\hbar\omega_{J_i}\cos\hat{\varphi}_i\right) \nonumber\\
	& \ \ \ \ \ \ \ -\hbar\omega_{J_8}
	\cos(\hat{\varphi}_5-\hat{\varphi}_4-\Theta_{\rm ex,L}), \\
	&\hat{H}_{67}=\sum_{i=6}^7 
	\left(4\hbar W_{ii} \hat{n}_i^2-\hbar\omega_{J_i}\cos\hat{\varphi}_i\right) \nonumber\\
	&\ \ \ \ \ \ \ -\hbar\omega_{J_9}\cos(\hat{\varphi}_7-\hat{\varphi}_6-\Theta_{\rm ex,R}), \\
	&\hat{H}_{\rm int} = \sum_{i=1}^3\sum_{j=4}^7 8\hbar W_{ij} \hat{n}_i\hat{n}_j
	+\sum_{i=4}^5\sum_{j=6}^7 8\hbar W_{ij} \hat{n}_i\hat{n}_j.
\end{align}
Here,
$\hat{H}_{123}$, $\hat{H}_{45}$, and $\hat{H}_{67}$ 
correspond to the qubit, DTC-L, and DTC-R subspaces 
, respectively,  
and $\hat{H}_{\rm int}$ is the interaction Hamiltonian between them. 

The qubit-subspace Hamiltonian 
$\hat{H}_{123}$ is expressed by $(2N+1)^3\times(2N+1)^3$ matrix. 
We diagonalize it using SciPy\cite{scipy} to obtain 
the eigenenergy matrix $\hat{e}_{123}$ 
and the energy eigenstate matrix $\hat{V}_{123}$ satisfiyng the following relation: 
\begin{align}
	\hat{V}_{123}^\dag\hat{H}_{123}\hat{V}_{123}=\hat{e}_{123}, 
\end{align}
where the diagonal components of 
$\hat{e}_{123}$
are the eigenenergies of $\hat{H}_{123}$ in the ascending order. 
We introduce the cutoff $N_{123}$
to restrict the matrix size of $\hat{e}_{123}$ and $\hat{V}_{123}$ to $N_{123}\times N_{123}$.  
Using $\hat{V}_{123}$, 
we can obtain 
$\hat{n}'_{i}$ for $i\in\{1,2,3\}$, 
which is the $\hat{n}_{i}$ represented by the $\hat{V}_{123}$ basis as follows: 
\begin{align}
	&\hat{n}'_i=\hat{V}^{\dag}_{123}\hat{n}_{i}\hat{V}_{123}\ {\rm for}\ i\in\{1,2,3\}. 
\end{align}
This procedure reduces the matrix size of $\hat{n}_i$ from 
$(2N+1)^3\times(2N+1)^3$ to $N_{123} \times N_{123}$.
In this work, we set $N_{123}$ to 120. 
Similarly, 
we reduce the matrix size of $\hat{n}_{i}$ for $i\in\{4,5,6,7\}$ as follows: 
\begin{align}
	&\hat{n}'_i=\hat{V}^{\dag}_{45}\hat{n}_{i}\hat{V}_{45}\ {\rm for}\ i\in\{4,5\}, \\
	&\hat{n}'_i=\hat{V}^{\dag}_{67}\hat{n}_{i}\hat{V}_{67}\ {\rm for}\ i\in\{6,7\}, 
\end{align}
where 
the eigenstate matrices
$\hat{V}_{45}$ and $\hat{V}_{67}$ 
satisfy the following equations:  
\begin{align}
	&\hat{V}_{45}^\dag \hat{H}_{45}\hat{V}_{45}=\hat{e}_{45}, \\ 
	&\hat{V}_{67}^\dag \hat{H}_{67}\hat{V}_{67}=\hat{e}_{67}. 
\end{align}
The diagonal components of the eigenenergy matrices 
$\hat{e}_{45}$ and $\hat{e}_{67}$
are the eigenenergies of $\hat{H}_{45}$ and $\hat{H}_{67}$, 
respectively, in the ascending order. 
Here, 
we introduce the cutoff $N_{45}$ and $N_{56}$
to restrict the matrix size of 
($\hat{e}_{45}$, $\hat{V}_{45}$) and ($\hat{e}_{67}$, $\hat{V}_{67}$)
to $N_{45}\times N_{45}$ and $N_{67}\times N_{67}$, respectively. 
In this work, we set both $N_{45}$ and $N_{67}$ to 25. 

Using them, 
we obtain the Hamiltonian with reduced matrix size as follows: 
\begin{align}
	&\hat{H}'
	=
	\hat{e}_{123}+\hat{e}_{45}+\hat{e}_{67}+\hat{H}'_{\rm int}, 
	\label{eq:Hprime}
	\\
	&\hat{H}'_{\rm int}
	=\sum_{i=1}^3\sum_{j=4}^7 8\hbar W_{ij} \hat{n}'_i\hat{n}'_j 
	+\sum_{i=4}^5\sum_{j=6}^7 8\hbar W_{ij} \hat{n}'_i\hat{n}'_j. 
\end{align}
The size of these matrices is $(N_{123}\times N_{45}\times N_{67})\times(N_{123}\times N_{45}\times N_{67})=75000\times75000$. 
By diagonalizing $\hat{H}'$, 
we can obtain the energy eigenvalues of the total system. 

We have set
the dimensions as 
$N_{123}=120$ and $N_{45}=N_{67}=25$ 
for sufficient convergence of eigenfrequencies with sub-kHz accuracy. 
By this method, 
the matrix size of the Hamiltonian for diagonalizations 
is greatly reduced 
from $(2N+1)^7\times(2N+1)^7=1801088541\times1801088541$ to $(N_{123}\times N_{45}\times N_{67})\times(N_{123}\times N_{45}\times N_{67})=75000\times75000$.

\subsection{Gate simulation}
We explain the gate simulation method. 
As an example, 
we cosider the case of CZ gate for $\rm{Q_1}$ and $\rm{Q_2}$. 

From the results of the diagonalization of the total Hamiltonian 
explained in the previous subsection, 
we identify the idle point. 
Fixing $\Theta_{\rm ex,R}$ to $\Theta_{\rm Id,R}$, 
we again diagonalize $\hat{H}'$ 
for $\Theta_{\rm ex,L}=\Theta_{n}=0.05n\pi\ (n=0,1,\cdots,20)$ 
to obtain the eigenenergy matrix $\hat{e}_{0}(\Theta_n)$ 
and the energy eigenstate matrix $\hat{V}_{0}(\Theta_n)$ satisfiyng the following relation: 
\begin{align}
	\hat{V}_{0}(\Theta_n)^\dag\hat{H}'(\Theta_n)\hat{V}_{0}(\Theta_n)=\hat{e}_{0}(\Theta_n), 
\end{align}
where the diagonal components of 
$\hat{e}_{0}(\Theta_n)$ are the eigenenergies of $\hat{H}'(\Theta_n)$ in the ascending order. 
We introduce the cutoff $N_{0}$
to restrict the matrix size of $\hat{e}_{0}(\Theta_n)$ 
and $\hat{V}_{0}(\Theta_n)$ to $N_{0}\times N_{0}$. 
In this work, we set $N_0$ to 1000. 
Using $\hat{V}_{0}(\Theta_n)$, 
we calculate the following matrices: 
\begin{align}
	&\hat{P}(\Theta_{n}) 
	= \left[\hat{V}_{123}\otimes\hat{V}_{45}(\Theta_{n})\otimes\hat{V}_{67}\right]
	   \hat{V}_0(\Theta_{n}), 
\end{align}
and express the Hamiltonian at the magnetic flux 
$\Theta_{\rm ex,L} \in \mathcal{R}_m=[\Theta_{m}-0.025\pi,\Theta_{m}+0.025\pi)$ as
\begin{align}
	\hat{H}''(\Theta_{\rm ex,L})&=\hat{e}_0(\Theta_{m}) \nonumber\\
	&+
	\dot{\Theta}_{\rm ex,L}
	\mathbf{t}_{\rm DL}^{\rm T}
	\hat{P}^{\dag}(\Theta_m) \hbar W \hat{\mathbf{n}}\hat{P}(\Theta_m) \nonumber\\
	&-\hbar\omega_{J_8}
	\hat{P}^{\dag}(\Theta_m)\hat{c}(\Theta_{\rm ex,L})\hat{P}(\Theta_m), \\
	\hat{c}(\Theta_{\rm ex,L})&=\cos(\hat{\varphi}_5-\hat{\varphi}_4-\Theta_{\rm ex,L}) \nonumber\\
	& -\cos(\hat{\varphi}_5-\hat{\varphi}_4-\Theta_{m}).
\end{align}
The matrix size of $\hat{H}''$ is $N_{0}\times N_{0}=1000\times1000$. 
When $\Theta_{\rm ex,L}(t)\in \mathcal{R}_m$, 
we calculate the time evolution of the state vector 
represented by the $\hat{P}_0(\Theta_{m})$ basis, 
using the above Hamitonian $\hat{H}''$ and QuTiP\cite{qutip1,qutip2}.
At the particular time 
when the range including $\Theta_{\rm ex,L}$ changes from 
$\mathcal{R}_m$ to $\mathcal{R}_{m+1}$ or to $\mathcal{R}_{m-1}$, 
we operate the following basis transformation matrices
\begin{align}
	&\hat{P}_{\rm inc}(\Theta_{m}) 
	= \hat{P}(\Theta_{m+1}) \hat{P}^{\dag}(\Theta_{m}), \\
	&\hat{P}_{\rm dec}(\Theta_{m}) 
	= \hat{P}(\Theta_{m-1}) \hat{P}^{\dag}(\Theta_{m}), 
\end{align}
respectively, on the state vector.

We have set the dimensions as 
$N_{123}=120$, $N_{45}=N_{67}=25$, and $N_{0}=1000$ 
for sufficient convergence of average fidelities with accuracy 
of the order of $10^{-5}$. 
By this method, 
the matrix size of the Hamiltonian for gate simulations 
is greatly reduced 
from $(2N+1)^7\times(2N+1)^7=1801088541\times1801088541$ to $N_{0}\times N_{0}=1000\times1000$.

\section{Isolated Two-Qubit Subsystems L and R}
\label{sec:IsolatedTwoQubitSubsystems}
\subsection{DTC architecture}
The Hamiltonian of the system $\mu\in\{{\rm L},{\rm R}\}$ 
is denoted by $\hat{H}^{\mu}$ 
(see Appendix \ref{sec:derivation} and Ref.\ \onlinecite{goto2022DTC} for the help of the derivation). 
Its eigenfrequencies and corresponding eigenstates 
are denoted by $\omega^{\rm L}_{Q_1,Q_2,C_4,C_5}\ \left(\omega^{\rm R}_{Q_2,Q_3,C_6,C_7}\right)$ 
and $\ket{Q_1,Q_2,C_4,C_5}_{\rm L}\ \left(\ket{Q_2,Q_3,C_6,C_7}_{\rm R}\right)$, respectively. 
We also use notations 
$\omega^{\rm L}_{Q_1,Q_2}\equiv\omega^{\rm L}_{Q_1,Q_2,0,0}\ 
\left(\omega^{\rm R}_{Q_2,Q_3}\equiv\omega^{\rm R}_{Q_2,Q_3,0,0}\right)$ and 
$\ket{Q_1,Q_2}_{\rm L}\equiv\ket{Q_1,Q_2,0,0}_{\rm L}\ 
\left(\ket{Q_2,Q_3}_{\rm R}\equiv\ket{Q_2,Q_3,0,0}_{\rm R}\right)$. 
Eigenfrequencies of the computational states in the system L (R), 
$\omega^{\rm L (R)}_{0,0}, \omega^{\rm L(R)}_{1,0}, \omega^{\rm L(R)}_{0,1}, \omega^{\rm L(R)}_{1,1}$, 
as functions of $\Theta_{\rm ex,L(R)}$ are shown by scatter plots 
in Fig. \ref{fig:DTCspectrumZZ}(a)[\ref{fig:DTCspectrumZZ}(b)]
where we have set $\omega_{0,0}^{\rm L(R)}$ to 0 (outside the graph scale). 

The $ZZ$ couplings, 
$\zeta_{12}$ in the system L and 
$\zeta_{23}$ in the system R 
are calculated as follows: 
\begin{align}
	\zeta_{12}
	&=
	\omega^{\rm L}_{1,1}
	-\omega^{\rm L}_{1,0}
	-\omega^{\rm L}_{0,1}
	+\omega^{\rm L}_{0,0}, 
	\label{eq:ZZ_twoqubitsL} \\
	\zeta_{23}
	&=\omega^{\rm R}_{1,1}
	- \omega^{\rm R}_{1,0}
	-\omega^{\rm R}_{0,1}
	+\omega^{\rm R}_{0,0}.
	\label{eq:ZZ_twoqubitsR}
\end{align}
These values should be as small as possible at the idle point 
to suppress conditional phase rotation which causes quantum crosstalk.
As shown in Fig.\ \ref{fig:DTCspectrumZZ}(c)[\ref{fig:DTCspectrumZZ}(d)] by the scatter plot, 
the minimum value of $|\zeta_{12(23)}|/(2\pi)$
is about $2(2)$ kHz at $\Theta_{\rm ex,L(R)}=0.65\pi(0.65\pi)\equiv\Theta_{\rm Id,L(R)}$. 
Hence, 
we can choose $\Theta_{\rm Id,L}$ and $\Theta_{\rm Id,R}$ 
as idle points of system L and R, respectively.

\subsection{STC architecture}
Similarly to the DTC architecture, 
we define $\hat{H}^{\rm L(R)}$. 
Its eigenfrequencies and corresponding eigenstates 
are denoted by 
$\omega^{\rm L}_{Q_1,Q_2,C_4}\ \left(\omega^{\rm R}_{Q_2,Q_3,C_5}\right)$ 
and $\ket{Q_1,Q_2,C_4}_{\rm L}\ \left(\ket{Q_2,Q_3,C_5}_{\rm R}\right)$, 
respectively. 
We also use notatations 
$\omega^{\rm L}_{Q_1,Q_2}\equiv\omega^{\rm L}_{Q_1,Q_2,0}\ 
\left(\omega^{\rm R}_{Q_2,Q_3}\equiv\omega^{\rm R}_{Q_2,Q_3,0}\right)$ and 
$\ket{Q_1,Q_2}_{\rm L}\equiv\ket{Q_1,Q_2,0}_{\rm L}\ 
\left(\ket{Q_2,Q_3}_{\rm R}\equiv\ket{Q_2,Q_3,0}_{\rm R}\right)$. 
Eigenfrequencies of the computational states in the system L (R), 
$\omega^{\rm L (R)}_{0,0}, \omega^{\rm L(R)}_{1,0}, \omega^{\rm L(R)}_{0,1}, \omega^{\rm L(R)}_{1,1}$, 
as functions of $\Theta_{\rm ex,L(R)}$ are shown by scatter plots 
in Fig. \ref{fig:STCspectrumZZ}(a)[\ref{fig:STCspectrumZZ}(b)] 
where we have set $\omega_{0,0}^{\rm L(R)}$ to 0 (outside the graph scale). 

The $ZZ$ couplings, 
$\zeta_{12}$ in the system L and 
$\zeta_{23}$ in the system R 
are calculated 
by Eq.\ (\ref{eq:ZZ_twoqubitsL}) and Eq.\ (\ref{eq:ZZ_twoqubitsR}), respectively. 
As shown in Fig.\ \ref{fig:STCspectrumZZ}(c)[\ref{fig:STCspectrumZZ}(d)] by the scatter plot, 
the minimum value of $|\zeta_{12(23)}|/(2\pi)$
is about $64(64)$ kHz at $\Theta_{\rm ex,L(R)}=0(0)\equiv\Theta_{\rm Id,L(R)}$. 
Hence, 
we choose $\Theta_{\rm Id,L}$ and $\Theta_{\rm Id,R}$ as idle points of system L and R, respectively. 
The residual couplings are summarized in Table\ \ref{tab:STCZZ}.
They are in good agreement with the result of Ref.\ \onlinecite{ETH}: about $-60$ kHz. 
Their magnitudes are roughly 30 times larger than the ones of the corresponding subsystems with the DTC. 

\section{Definition of $\hat{U}'$}
\label{sec:Uprime}
Here we explain the derivation of $\hat{U}'$ in Eq.\ (\ref{eq:fidelity}) for average fidelity.
\subsection{Three-qubit system}
Using $\widetilde{\ket{Q_1',Q_2',Q_3'}}$, 
which is the resultant state of a gate operation on $\ket{Q_1',Q_2',Q_3'}$ ($Q_1',Q_2',Q_3'\in\{0,1\}$), 
we define $\hat{u}'$ as follows: 
\begin{align}
	&u'_{4Q_1+2Q_2+Q_3,4Q_1'+2Q_2'+Q_3'}
	\nonumber\\
	&=
	\frac{\left(\braket{0,0,0|\widetilde{0,0,0}}\right)^*}{\left|\braket{0,0,0|\widetilde{0,0,0}}\right|}
	\braket{Q_1,Q_2,Q_3|\widetilde{Q_1',Q_2',Q_3'}}, 
	\label{eq:uprime}
\end{align}
where 
we have choosen the overall phase factor, 
$\left(\braket{0,0,0|\widetilde{0,0,0}}\right)^*/\left|\braket{0,0,0|\widetilde{0,0,0}}\right|$,
such that $u'_{0,0}$ is equal to $|u'_{0,0}|$. 
Moreover, 
to account for the local single-qubit phase, 
we define $\hat{U}'$ as
\begin{align}
	&\hat{U}'= \hat{P}_{Z}(\psi_1,\psi_2,\psi_3) \hat{u}' \hat{P}_{Z}^{\dag}(\psi'_1,\psi'_2,\psi'_3), \\
	&\hat{P}_{Z}(\psi_1,\psi_2,\psi_3) = \nonumber\\
	& \ \ \ \ \ \ \ 
	\begin{pmatrix}
			  1 & 0  \\
 			  0 & e^{-i\psi_{1}}  \\ 
	\end{pmatrix}
	\otimes
	\begin{pmatrix}
			  1 & 0  \\
 			  0 & e^{-i\psi_{2}}  \\ 
	\end{pmatrix}
	\otimes
	\begin{pmatrix}
			  1 & 0  \\
 			  0 & e^{-i\psi_{3}}  \\ 
	\end{pmatrix}.
	\label{eq:Rz3}
\end{align}
Phase rotation angles $\psi_i$ and $\psi'_i$ 
should be chosen appropriately 
for each ideal gate operation $\hat{U}_{\rm id}$ (see Appendix \ref{sec:cariburation}).

\subsection{Two-qubit systems}
Similarly to Eq.\ (\ref{eq:uprime}), 
we define $\hat{u}'$ as follows: 
\begin{align}
	&u'_{2Q_1+Q_2,2Q_1'+Q_2'}
	\nonumber\\
	&=
	\frac{\left({}_{\rm L(R)}\braket{0,0|\widetilde{0,0}}_{\rm L(R)}\right)^*}
	{\left|{}_{\rm L(R)}\braket{0,0|\widetilde{0,0}}_{\rm L(R)}\right|}
	{}_{\rm L(R)}\braket{Q_1,Q_2|\widetilde{Q_1',Q_2'}}_{\rm L(R)}.   
	\label{eq:up2}
\end{align}
Moreover, 
to account for the local single-qubit phase, 
we define $\hat{U}'$ as follows: 
\begin{align}
	&\hat{U}'= \hat{P}_{Z}(\psi_1,\psi_2) \hat{u}' \hat{P}_{Z}^{\dag}(\psi'_1,\psi'_2), \\
	&\hat{P}_{Z}(\psi_1,\psi_2)=
	\begin{pmatrix}
			  1 & 0  \\
 			  0 & e^{-i\psi_{1}}  \\ 
	\end{pmatrix}
	\otimes
	\begin{pmatrix}
			  1 & 0  \\
 			  0 & e^{-i\psi_{2}}  \\ 
	\end{pmatrix}.
	\label{eq:Rz2}
\end{align}
Phase rotation angles $\psi_i$ and $\psi'_i$ should be chosen appropriately 
for each ideal gate operation $\hat{U}_{\rm id}$ (see Appendix \ref{sec:cariburation}).

\begin{figure*}[t]
        \centering
        \hspace{-1cm}
        \includegraphics[width=0.95\textwidth]{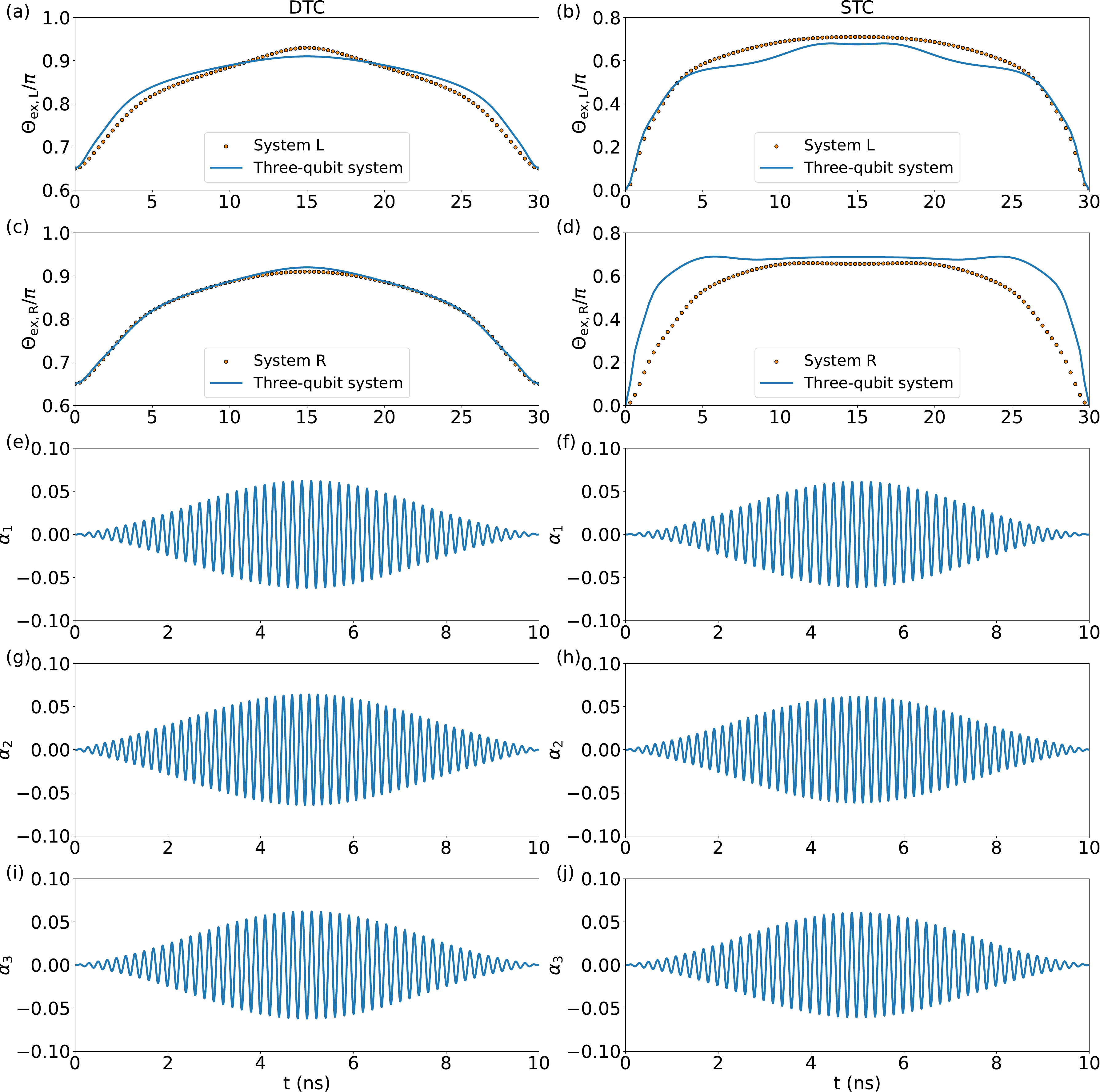}
  	   	\caption{
  	   		Optimized gate pulses. 
  	   		The left and right column
  	   		correspond to the
  	   		DTC and STC architectures, respectively. 
			(a) and (b): optimized $\Theta_{\rm ex,L}(t)$.
			(c) and (d): optimized $\Theta_{\rm ex,R}(t)$. 
			In (a)--(d), 
			solid curves and scatter plots 
			represent the pulses in three-qubit systems and two-qubit subsystems, respectively. 
			(e) and (f): optimized $\alpha_{1}(t)$. 
			(g) and (h): optimized $\alpha_{2}(t)$. 
			(i) and (j): optimized $\alpha_{3}(t)$. 
			}
  	   \label{fig:pulse}
\end{figure*}
\section{Phase caliburation}
\label{sec:cariburation}
\subsection{CZ gate}
As an example, 
we consider a CZ gate for $({\rm Q_1}, {\rm Q_2})$. 
In this case, 
we define the above angles as
\begin{align}
	\psi_{1}&=\arg(u'_{44}), \\
	\psi_{2}&=\arg(u'_{22}), \\
	\psi_{3}&=\arg(u'_{11}), \\
	\psi'_{1}&=\psi'_{2}=\psi'_{3}=0. 
\end{align}
This is because when leakage errors, 
undesired transition, and undesired conditional phase rotation
do not exist,  
the matrix form of $\hat{u}'$ can be written as follows:   
\begin{align}
	\hat{u}'&=
	\begin{pmatrix}
			  1 & \ & \ &\ \\
 			  \ & e^{i\phi_{010}} &\ &\ \\ 
 			  \ &\ &e^{i\phi_{100}} &\ \\
 			   \ &\ &\ &-e^{i(\phi_{010}+\phi_{100})} \\
		\end{pmatrix}
	\otimes
	\begin{pmatrix}
			  1 & 0  \\
 			  0 & e^{i\phi_{001}}  \\ 
	\end{pmatrix}\nonumber\\
	&=
	\hat{P}_{Z}^\dag(\phi_{100},\phi_{010},\phi_{001})
	\hat{U}_{CZ,12},
\end{align}
where $\phi_{100}=\arg(u'_{44}),\phi_{010}=\arg(u'_{22})$, and $\phi_{001}=\arg(u'_{11})$. 
Similar caliburations are valid for the CZ gate for $({\rm Q_2}, {\rm Q_3})$ 
and the ones in the two-qubit subsystems.

\subsection{$\pi/2$ pulse}
Next, as an example, 
we consider a $\pi/2$ pulse for $\rm{Q_1}$. 
In this case, 
we define the angles as  
\begin{align}
	\psi_{1}&=\arg(u'_{44})-\arg(-u'_{04}), \\  
	\psi_{2}&=\arg(u'_{22}), \\
	\psi_{3}&=\arg(u'_{11}), \\
	\psi'_{1}&=-\arg(iu'_{04}), \\
	\psi'_{2}&=\psi'_{3}=0. 
\end{align}
This is because 
when leakage errors, 
undesired transition, and undesired conditional phase rotation
do not exist,  
the matrix form of $\hat{u}'$ can be written as follows:   
\begin{align}
\hat{u}'&=
	\frac{1}{\sqrt{2}}
	\begin{pmatrix}
			  1 & -ie^{i\phi_{A}}  \\
 			  -ie^{i(\phi_{100}-\phi_A)} & e^{i\phi_{100}}  \\ 
	\end{pmatrix}
	\otimes
	\begin{pmatrix}
			  1 & 0  \\
 			  0 & e^{i\phi_{010}}  \\ 
	\end{pmatrix}
	\otimes
	\begin{pmatrix}
			  1 & 0  \\
 			  0 & e^{i\phi_{001}}  \\ 
	\end{pmatrix}\nonumber\\
	&=
	\hat{P}_{Z}^\dag(\phi_{100}-\phi_{A},\phi_{010},\phi_{001})
	\hat{U}_{\pi/2,\{1\}}
	\hat{P}_{Z}(-\phi_{A},0,0),  
\end{align}
where $\phi_{A}=\arg(iu'_{04})$. 
Similar caliburations are valid for the $\pi/2$ pulses
for ${\rm Q_2}$ and ${\rm Q_3}$.

\section{Pulse optimization}
\subsection{CZ gate}
\label{sec:CZ_Optimization}
We design the dc flux pulses for implementing CZ gates
by the known technique 
to suppress leakage errors\cite{Martinis,goto2022DTC,KG2023}. 
In the 3rd-order design,
$\Theta_{\rm ex,\mu}(t)$ 
has the following form:
\begin{align}
	&\theta_{\rm L(R)}(t)
	={f_{\rm L(R)}[\Theta_{\rm Id,L(R)}]} \nonumber\\
	&{+\frac{f_{\rm L(R)}[\Theta_{\rm ex,L(R)}(t)]-f_{\rm L(R)}[\Theta_{\rm Id,L(R)}]}{2}} \nonumber\\
	& \ \ \ \ \  \times
	\sum_{n=1}^3\lambda_n
	{\left(
	1-\cos\frac{2n\pi t}{T}
	\right)},   
\end{align}
where the hyper parameter $\lambda_n$ satisfies $\lambda_1+\lambda_3=1$ 
and $f_{\rm L(R)}$ is a numerical function of $\Theta_{\rm ex,L(R)}$, 
which can be determined from 
an aditional hyper parameter $\Theta_{\rm P,L(R)}$ corresponding to a pulse peak 
and the spectrum data 
\{Fig.\ \ref{fig:DTCspectrumZZ}(a)[\ref{fig:DTCspectrumZZ}(b)]
and Fig.\ \ref{fig:STCspectrumZZ}(a)[\ref{fig:STCspectrumZZ}(b)]
for the system with DTCs and STCs, respectively\}
(see Ref.\ \onlinecite{Martinis} and Appendix C of Ref.\ \onlinecite{goto2022DTC} for the details). 
Applying the above $\Theta_{\rm ex, L(R)}(t)$ , 
we calculate infidelity 1-$\bar{F}_{CZ,12(23)}$ as a cost function, 
and optimize $\lambda_n$ 
by using 
python version of the L-BFGS-B optimizer of the optimparallel package\cite{optimparallel1,optimparallel2}. 
We iterate this optimization for different $\Theta_{\rm P,L(R)}\in[\Theta_{\rm Id,L(R)},\pi]$ in increments of 0.01$\pi$. 
We obtain the optimal pulse as the one with the lowest infidelity [see Figs.\ \ref{fig:pulse}(a)-\ref{fig:pulse}(d) for their shapes].

\subsection{$\pi/2$ pulse}
\label{sec:piover2_Optimization}
It is known that 
applying a non-zero $\alpha_i$ modulated at the 
qubit frequency $\omega_{{\rm Id},i}$ of 
${\rm Q}_i$ at the idle point induces a Rabi ocillation in the ${\rm Q}_i$ subspace.
To supress the nonadiabatic error, 
we asuume the following DRAG-like pulse shape\cite{DRAG}:  
\begin{align}
	\alpha_{i} &= A_{i}(t)\sin(\omega_{{\rm Id},i}t) + B_{i}(t)\cos(\omega_{\rm{Id},i}t), \\
	A_i &= a_i \frac{1}{2}{\left(1-\cos\frac{2\pi t}{T}\right)}, \\
	B_i &= b_i \frac{\pi}{T} \frac{1}{\eta_i} \sin\frac{2\pi t}{T},
\end{align}
where $\eta_i$ is the anharmonicity of the ${\rm Q}_i$ at the idle point, 
and $T$ is the gate time. 
The values $a_i$ and $b_i$ are hyperparameters that determine the pulse shape.
Applying the above $\alpha_i$, 
we calculate infidelity 1-$\bar{F}_{\pi/2,\{i\}}$ as a cost function, 
and optimize $a_i$ and $b_i$ 
by using 
python version of the L-BFGS-B optimizer of the optimparallel package\cite{optimparallel1,optimparallel2}. 
We obtain the optimal pulse as the one with the lowest infidelity [see Figs.\ \ref{fig:pulse}(e)-\ref{fig:pulse}(j) for their shapes].

\section{Error analysis}
\label{appendix:errors}
In this sections, 
we demonstrate that  
the STC architecture exhibits more pronounced errors 
in the ${\rm Q}_1$-${\rm Q}_3$ subspace 
compared to the DTC architecture.

\begin{figure}[t]
	\centering
	\includegraphics[width=0.5\textwidth]{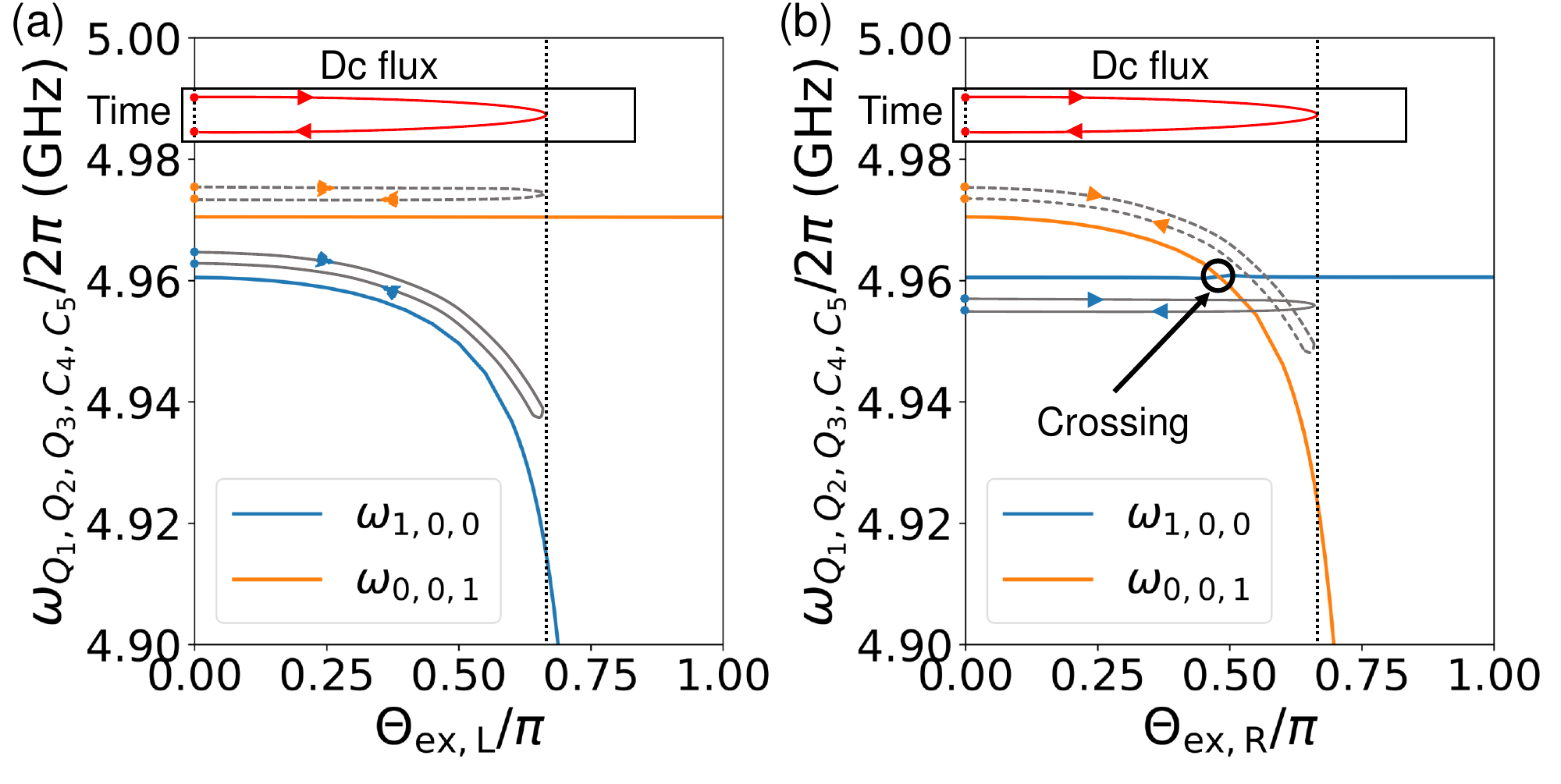}
	\caption{
	Qubit frequencies during the CZ gate. 
	(a) and (b) show qubit frequencies, $\omega_{1,0,0}$ and $\omega_{0,0,1}$, 
	as functions of $\Theta_{\rm ex,L}$ and $\Theta_{\rm ex,R}$, respectively. 
	The inset figures show typical flux-pulse shapes 
	[$\Theta_{\rm ex,L}(t)$ and $\Theta_{\rm ex,R}(t)$]
	for implementing CZ gates.
	The vertical dotted lines indicate the pulse peaks.
	The gray-solid lines and gray-dashed ones with arrows indicate 
	trajectories of $\omega_{1,0,0}$ and $\omega_{0,0,1}$, respectively. 
	Their trajectories are crossed only in (b). 
	Note that
	$\Theta_{\rm ex,R}$ in (a) and $\Theta_{\rm ex,L}$ in (b) 
	are fixed to the idle points.	
	}
\label{fig:STCcross}
\end{figure}
\subsection{Microwave crosstalk during single-qubit gate operation}
\label{sec:Rabi}
We first examine 
typical microwave crosstalk: 
undesired Rabi oscillation in the ${\rm Q}_{i}$ subspace 
by the microwave pulse $\alpha_{j}$ $(i\neq j)$. 
We denote its rotation angle as $\theta_{ji}$ 
and evaluate it from $\hat{U}'$ 
implemented by optimized microwave pulses $\alpha_j$. 

We find that, in the DTC architecture, 
$|\theta_{13}|\sim|\theta_{31}|\sim10^{-3}\pi$ 
and the other rotation angles 
are equal to or smaller than 
$10^{-5}\pi$.  
They are small enough to implement
$\pi/2$ pulses with a fidelity over 99.99\%. 

On the other hand, 
in the STC architecture, 
while the most  $|\theta_{ij}|$ are the same order as the DTC,   
only $|\theta_{13}|$ and $|\theta_{31}|$ are roughly 10 times larger. 
This may be caused by a relatively large $g_{13}$ of the STC architecture, 
as discussed in Sec.\ \ref{sec:Discussion}.  
These results imply that the DTCs are more robust against microwave crosstalk than the STCs.

\subsection{Transition between $\ket{1,Q_2,0}$ and $\ket{0,Q_2,1}$ during CZ-gate operation}
\label{sec:transition}
Next, we examine the errors during the CZ-gate operation 
from $\hat{U}'$ implemented by the flux pulse $\Theta_{\rm ex,L(R)}$. 

As a result, 
we find that 
all the off-diagonal elements of $\hat{U}'$ have 
an absolute value smaller than $10^{-2}$ 
in the DTC architecture. 
In other words, 
undesired transitions are almost negligible during a CZ gate in the DTC architecture, 
leading 
to the average gate fidelity of 99.99\%.

On the other hand, 
in the STC architecture, 
the off-diagonal elements of $\hat{U}'$ corresponding to 
$\braket{1,Q_2,0|\widetilde{0,Q_2,1}}$ 
and $\braket{0,Q_2,1|\widetilde{1,Q_2,0}}$ 
are larger than $10^{-2}$. 
This result indicates that
the SWAP-like transitions in the $\rm{Q_1}$-$\rm{Q_3}$ subspace occur during a CZ-gate operation.
This leads to the gate-fidelity degradation. 
Furthermore, we found that 
the off-diagonal elements becomes larger for
a CZ-gate operation on $(\rm{Q_2}, \rm{Q_3})$ 
than on 
$(\rm{Q_1}, \rm{Q_2})$.
This can be explained by 
the crossing point 
of $\omega_{1,Q_2,0}$ and $\omega_{0,Q_2,1}$ 
during a CZ-gate operation on $(\rm{Q_2}, \rm{Q_3})$,
as shown in Fig. \ref{fig:STCcross}. 
The transitions corresponding to the off-diagonal elements 
lead to the result $\bar{F}_{CZ,23}<\bar{F}_{CZ,12}$ in Table\ \ref{tab:STCZZ}. 
These serious transitions 
also may be caused by a large $g_{13}$ discussed in Sec.\ \ref{sec:Discussion}.

\begin{figure*}[t]
\includegraphics[width=\textwidth]{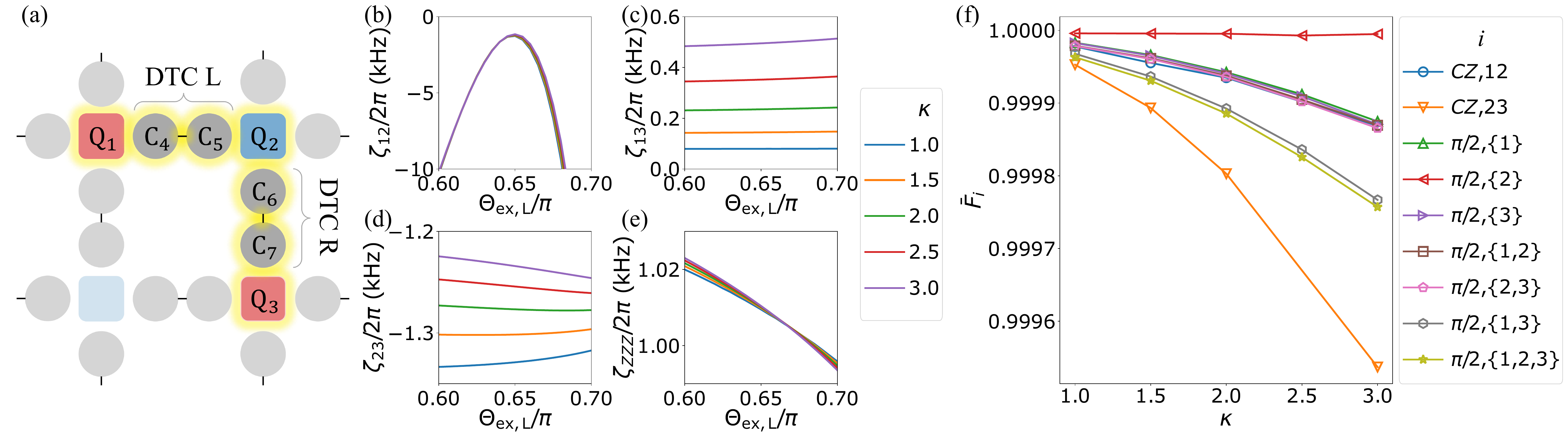}
\caption{
(a) A three-qubit system of diagonal structure (highlighted part). 
(b)-(e) $ZZ$ and $ZZZ$ couplings around the idle point as functions of $\Theta_{\rm ex, L}. $
Here, $\Theta_{\rm ex, R}$ is fixed to $\Theta_{\rm Id, R}$. 
(f) The average gate fidelities of 30-ns CZ gates and 10-ns $\pi/2$ pulses, implemented by optimized pulses, as functions of $\kappa$. 
}
\label{fig:diagonal}
\end{figure*} 
\section{Diagonal setting}
\label{appendix:diagonal}
In the main text, we considered the three-qubit systems in which the NNN qubits are located on the horizontal line like Fig.\ \ref{fig:redblue}(a). 
To scale up a superconducting quantum computer, 
the diagonal structure such as the highlighted part of Fig.\ \ref{fig:diagonal}(a) is also important.
Since the distances $\rm{Q_1}$-$\rm{Q_3}$, $\rm{Q_{1(3)}}$-DTC R(L), 
and DTC L-DTC R in this structure tend to be closer 
than the ones of the horizontal structure, 
the parasitic capacitances between them tend to increase. 
Therefore, it is necessary to clarify how the performance of the three-qubit system changes 
in response to the increases of the parasitic capacitances.
Here, we investigate the case where the above parasitic capacitances, 
$C_{ij}=C_{ji}\ {\rm for}\ ij\in\{13,16,17,34,35,46,47,56,57\}$, 
are increased by a factor of 
$\kappa\in{[1.0,3.0]}$ 
as follows: 
\begin{align}
	&C_{ij(ji)}\rightarrow \kappa C_{ij(ji)}. 
	\label{eq:kappa}
\end{align}

Figures \ref{fig:diagonal}(b)-\ref{fig:diagonal}(e) 
show the $ZZ$ and $ZZZ$ couplings around the idle point. 
We find that 
the residual couplings do not increase by more than 1 kHz 
up to $\kappa=3$. 
It means that
the ability of the residual coupling suppression of the DTC 
is highly robust against the increase of the parasitic capacitances. 

Figure \ref{fig:diagonal}(f) 
shows the average gate fidelities of 30-ns CZ gates and 10-ns $\pi/2$ pulses, 
implemented by optimized pulses, as functions of $\kappa$. 
The fidelities
tend to decrease as $\kappa$ 
increases, in particular, for 
$\bar{F}_{CZ,23}$, $\bar{F}_{\frac{\pi}{2},\{1,3\}}$ and $\bar{F}_{\frac{\pi}{2},\{1,2,3\}}$.  
This may be due to the effect of the crossing point and microwave crosstalk 
discussed in Appendix \ref{appendix:errors}.
However, when $\kappa\lesssim 1.5$, 
all the gates can be implemented 
with fidelities over 99.99\% in the 4-digit precision.
Such parasitic capacitances 
may be experimentally feasible. 
Therefore, 
from these results, 
we expect that the DTC architecture can 
be extended to multi-qubit systems while keeping its high performance.

\end{document}